\documentclass[a4paper,12pt,final]{article}
\usepackage{etex}

\usepackage[sc]{mathpazo}

\usepackage[table]{xcolor}

\usepackage{threeparttable}
\usepackage{standalone}
\usepackage{booktabs, dcolumn}
\usepackage{booktabs}

\usepackage{afterpage}

\usepackage{rotating}
\usepackage{tikz}

\usepackage{latexsym}
\usepackage{amsfonts}
\usepackage{amssymb}
\usepackage{amsthm}
\usepackage{amsmath}
\usepackage[mathscr]{eucal}
\usepackage{multicol}
\usepackage{multirow}
\usepackage{setspace}
\usepackage{graphicx}
\usepackage{pdfpages}
\usepackage{pdflscape}

\usepackage{anyfontsize}
\usepackage{t1enc}

\usepackage{caption}
\usepackage{subcaption}
\DeclareCaptionFont{captiofont}{\fontsize{9pt}{10.8pt}\selectfont}
\usepackage{verbatim}
\definecolor{myred}{rgb}{0.89412, 0.10196, 0.10980}
\definecolor{myblue}{rgb}{0.21569, 0.49412, 0.72157}
\definecolor{mygreen}{rgb}{0.30196, 0.68627, 0.29020}
\definecolor{mygray}{rgb}{0.90, 0.90, 0.90}

\usepackage[plainpages=false,
breaklinks=true,
colorlinks=true,
urlcolor=gray,
citecolor=blue, %blue , RoyalBlue
linkcolor=red, %red, %black BrickRed
bookmarks=true,
bookmarksopen=true,
bookmarksopenlevel=1,
pdfstartview=FitH,
pdfview=FitH]{hyperref}

\usepackage{color,hyperref}

\usepackage[round]{natbib}
\usepackage[lmargin=1in, rmargin=1in, tmargin=1in, bmargin=1in, paperwidth=220mm, paperheight=290mm]{geometry}

\usepackage{booktabs}
\usepackage{tabulary}
\usepackage{tabularx}
\usepackage{multirow}
\usepackage{array}

\usepackage{enumerate}

\usepackage{sectsty}
\usepackage{lipsum}
\sectionfont{\centering}

\usepackage{tocloft}
\usepackage{titlesec}
\titleformat{\section}
{\large\bfseries}{\thesection}{0.5em}{}

\renewcommand\thesection{\arabic{section}}

\renewcommand\thesubsection{\thesection.\arabic{subsection}}
\titleformat{\subsection}
{\normalfont\itshape}{\thesubsection}{0.5em}{}

\renewcommand\thesubsubsection{\thesection.\arabic{subsection}.\arabic{subsubsection}}
\titleformat{\subsubsection}{\normalfont\itshape}{\thesubsubsection}{0.5em}{}

\usepackage{setspace}

\titlespacing\section{0pt}{10pt plus 4pt minus 2pt}{5pt plus 2pt minus 2pt}
\titlespacing\subsection{0pt}{10pt plus 4pt minus 2pt}{0pt plus 2pt minus 2pt}
\titlespacing\subsubsection{0pt}{10pt plus 4pt minus 2pt}{0pt plus 2pt minus 2pt}

\providecommand{\keywords}[1]{\textbf{Keywords:}  #1}
\providecommand{\JEL}[1]{\textbf{JEL:}  #1}

\makeatletter
\newcommand*{\myfnsymbolsingle}[1]{%
\ensuremath{%
\ifcase#1% 0
\or % 1
*%
\or % 2
\dagger
\or % 3
\ddagger
\or % 4
\mathsection
\or % 5
\mathparagraph
\else % >= 6
\@ctrerr
\fi
}%
}
\makeatother

\usepackage{alphalph}
\newalphalph{\myfnsymbolmult}[mult]{\myfnsymbolsingle}{}

\renewcommand*{\thefootnote}{%
\myfnsymbolmult{\value{footnote}}%
}

\makeatletter
\def\@xfootnote[#1]{%
\protected@xdef\@thefnmark{#1}%
\@footnotemark\@footnotetext}
\makeatother

\usepackage{bbding}         % \FiveStarOpen

\usepackage[symbol]{footmisc}

\usepackage{chngcntr}

\usepackage{scrwfile}
\TOCclone[\contentsname]{toc}{atoc}

\AfterTOCHead[toc]{%
}
\AfterTOCHead[atoc]{%
\edef\maintocdepth{\the\value{tocdepth}}%
\value{tocdepth}=-10000\relax%
}

\newcommand{\chapquote}[3]{\begin{quotation} \textit{#1} \end{quotation} \begin{flushright} - #2, \textit{#3}\end{flushright} }

\usepackage{blindtext}

\usepackage{algorithm2e}

\begin{document}

\setcounter{footnote}{0}

\newpage

\title{\Large \setcounter{footnote}{2}Learning the Probability Distributions of Day-Ahead Electricity Prices\thanks{We are grateful to the editor, Lance Bachmeier, and two anonymous referees for their useful comments and suggestions, which have greatly improved the paper. We are grateful to Rafa\l{} Weron, Florian Ziel, Wolfgang Hardle, Arkadiusz Lipiecki, Lukas Vacha, Frantisek Cech, and the participants at various conferences and research seminars for their many useful comments, suggestions, and discussions. We gratefully acknowledge the support received from the Czech Science Foundation under the 24-11555S project. We provide the computational package \texttt{DistrNNEnegry.jl} in \textsf{JULIA} at \url{https://github.com/luboshanus/DistrNNEnergy.jl}, which will allow one to use our time series data measures.}
\vspace{20pt}}

\author{\setcounter{footnote}{6}Lubo\v{s} Hanus\thanks{Institute of Economic Studies, Charles University, Opletalova 26, 110 00, Prague, CR and Institute of Information Theory and Automation, Academy of Sciences of the Czech Republic, Pod Vodarenskou Vezi 4, 18200, Prague, Czech Republic. E-mail: \url{hanusl@utia.cas.cz}} \\
{\small\textit{Charles University and}} \\
{\small\textit{Czech Academy of Sciences}}
\and
\and
\setcounter{footnote}{0}Jozef Barun\'{i}k\thanks{Institute of Economic Studies, Charles University, Opletalova 26, 110 00, Prague, CR and Institute of Information Theory and Automation, Czech Academy of Sciences , Pod Vodarenskou Vezi 4, 18200, Prague, Czech Republic. E-mail: \url{barunik@utia.cas.cz }\,\, Web: \href{https://barunik.github.io/}{barunik.github.io}} \\ {\small\textit{Charles University and}} \\
{\small\textit{Czech Academy of Sciences}}}

\date{\today}
% \date{\hspace{2em}}

\maketitle

\begin{abstract}

We propose a novel machine learning approach for probabilistic forecasting of hourly day-ahead electricity prices. In contrast with the recent advances in data-rich probabilistic forecasting, which approximates distributions with few features (such as moments), our method is nonparametric and selects the distribution from all possible empirical distributions learned from the input data without the need for limiting assumptions. The model that we propose is a multioutput neural network that accounts for the temporal dynamics of the probabilities and controls for monotonicity using a penalty. Such a distributional neural network can precisely learn complex patterns from many relevant variables that affect electricity prices. We illustrate the capacity of the developed method on German hourly day-ahead electricity prices and predict their probability distribution via many variables, doing so more accurately than the state-of-the-art benchmarks can, thus revealing new valuable information in the data.
\vspace{10pt}

\keywords{Distributional forecasting, deep learning, probabilistic, electricity, energy time series}

\JEL{C45, C53, E17, E37}
\end{abstract}

\renewcommand{\thefootnote}{\arabic{footnote}}
\setcounter{footnote}{0}

% \linespread{1.0}
% \onehalfspacing
\doublespacing

\newpage

\section{Introduction}
\label{sec:intro}

\chapquote{``We will make electricity so cheap that only the rich will burn candles."}{Thomas A. Edison}{1880}

Electricity is crucial for modern economic and social activities. However, accurately forecasting electricity prices is inherently challenging because of their complex, nonlinear dynamics, which are influenced by demand--supply interactions and external market conditions. The increasing integration of intermittent renewable energy sources has also intensified the uncertainty exhibited by electricity prices, rendering traditional point forecasts inadequate for effectively making decisions, managing risks, trading and hedging. Instead, accurate probabilistic forecasting has become an indispensable process for energy producers, retailers, and traders, who require forecasts that explicitly represent uncertainty \citep{maciejowska2020assessing, bunn2016analysis}.

The literature addressing these urgent needs has focused primarily on statistical techniques, such as quantile regression \citep{uniejewski2021regularized,nitka2023combining,lipiecki2024postprocessing}, interval forecasting and conformal predictions \citep{xu2024novel,KATH2021777}, ensemble methods \citep{zhang2021day}, and hybrid deterministic--probabilistic approaches \citep{marcjasz2020probabilistic}. Recently, machine learning methods, particularly deep learning techniques, have gained prominence due to their ability to handle large amounts of data and identify complex patterns \citep{nowotarski2018recent,jkedrzejewski2022electricity,WeronMarcjasz:2023aa}. Specifically, models employing autoregressive recurrent neural networks \citep{salinas2020deepar}, nonlinear autoregressive neural networks \citep{marcjasz2020probabilistic}, and deep recurrent networks \citep{klein2023deep} have been proposed and extensively investigated in this context. Most recently, \cite{mashlakov2021assessing,WeronMarcjasz:2023aa} proposed deep learning models that enable multivariate forecasting in a context representing the current state-of-the-art capabilities. 

However, the current studies still struggle to accurately represent price uncertainty, which limits the guidance they can offer decision-makers for numerous reasons. Many approaches rely on restrictive model assumptions. For example, machine learning models learn the parameters of assumed distributions \citep{mashlakov2021assessing,WeronMarcjasz:2023aa} or target specific distribution characteristics, such as particular quantiles or moments, rather than modeling full distributions. Such strong parametric assumptions can lead to biased or misspecified predictive densities, particularly in light of the heavy-tailed and skewed price behaviors that are observed in practice. These models also tend to be heavily parameterized \citep{grothe2023point}, which reduces their ability to adapt to the rapidly changing, nonlinear dynamics of electricity markets. Furthermore, many traditional models integrate only exogenous drivers (e.g., weather forecasts and fuel prices) in a limited manner \citep{he2020end,memarzadeh2021short,jiang2024probabilistic}. Important effects, such as seasonal effects and regulatory changes, are often ignored or difficult to incorporate \citep{he2020end,jiang2024probabilistic}. Additionally, some techniques fail to exploit the complex nonlinear relationships contained in the given data \citep{uniejewski2021regularized,xu2024novel}, and the cutting-edge deep learning models often have an enormous computational complexity \citep{banitalebi2021intelligent,WeronMarcjasz:2023aa}, resulting in long training times.

These gaps highlight the need for more flexible approaches that can model full price distributions, leverage rich information, and remain tractable for operational use. In this paper, we address the question of how probability distributions can be learned from and forecasted using data without introducing confounding distributional assumptions. We do this by proposing a novel nonparametric distributional neural network (DistrNN) model with intrinsic distribution dynamics and a potentially large number of external variables. This approach makes several key contributions to the literature.
\begin{itemize}
\item A DistrNN framework forecasts entire probability distributions for hourly day-ahead electricity prices \textit{without restrictive assumptions about the underlying distributions.}
\item The model uniquely captures the \textit{intrinsic probability dynamics of price data}, effectively modeling nonlinear relationships.
\item Despite its complexity, the model remains \textit{computationally feasible and efficient}.
\item The model leverages \textit{extensive, high-dimensional data}, including historical prices, load forecasts, and external variables such as fuel and emission allowance prices.
\end{itemize}

Essentially, our proposed DistrNN approach uses high-dimensional input features to approximate the full empirical distribution of future prices without confounding distribution assumptions. This data-driven model captures the intrinsic dynamics of the distribution at all probability levels, enabling the model to reflect the evolution trend of price uncertainty over time. To our knowledge, this is the first framework to explicitly model this type of distribution in a fully nonparametric deep learning setting.

We empirically validate our model on German hourly day-ahead electricity price data incorporating 252 features, including lagged prices, loads and external market variables such as EU allowance prices and fuel costs. By benchmarking against prominent models, such as naive forecasting techniques, quantile regression averaging and quantile regression committee machines \citep{Nowotarski:2015aa,marcjasz2020probabilistic,uniejewski2021regularized,lipiecki2024postprocessing}, as well as the state-of-the-art distributional deep neural network (DDNN) \citep{WeronMarcjasz:2023aa}, we demonstrate that our method achieves good accuracy and reliability in terms of forecasting the entire distribution. While our method does not universally outperform all benchmarks, it significantly improves the forecasts produced during certain periods and therefore provides a useful complement to the existing methods, offering a nonparametric, assumption-light, data-rich approach coupled with an efficient computational framework that advances the probabilistic electricity price forecasting field. We also shed light on neural network models, showing that they provide different and complementary information.

\section{Probabilistic forecasting via a DistrNN} 
\label{sec:methods}

Consider a time series consisting of hourly day-ahead electricity prices $p_{t,h}$ collected over $t=1\ldots,T$ days and $h=1,\ldots,24$ hours. The main objective is to approximate the conditional cumulative distribution function (CDF) $F(p_{t,h}|\boldsymbol{z}_{t-1})$ as closely as possible and use it to make a $1$-step-ahead probabilistic forecast at time $t-1$ with a set of predictors $\boldsymbol{z}_{t-1}$ containing past values of $p_{t,h}$, past probability values, and past values of a possibly large set of exogenous observable variables $x_t$ that are relevant to prices. 

The main goal is then to approximate a collection of conditional probabilities that correspond to the empirical quantiles, such as 
\begin{equation}
\Big\{F(q_h^{\alpha_1}),\ldots,F(q_h^{\alpha_k})\Big\}=\Big\{ \Pr\Big(p_{t,h}\le q_h^{\alpha_1}|\boldsymbol{z}_{t-1}\Big), \ldots,\Pr\Big(p_{t,h}\le q_h^{\alpha_k}|\boldsymbol{z}_{t-1}\Big) \Big\}
\end{equation}
for the $k$ regularly spaced probabilities $0<\alpha_1<\alpha_2<\ldots<\alpha_k<1$. A convenient way to estimate such quantities is distribution regression. \cite{foresi1995conditional} noted that several binary regression models serve as good partial descriptors of the conditional distribution and proposed a distributional regression model
with (monotonically increasing) functions including logit, probit, linear, and log-log functions. \cite{chernozhukov2013inference} further considered a continuum of binary regression models, and \cite{barunik2019simple} suggested binding the coefficients of predictors in an ordered logit model via their smooth dependence on the corresponding probability levels.

\subsection{DistrNN}

Such probabilistic predictions are highly dependent on the parameterization of the developed model and quickly become infeasible as the number of covariates increases. This motivates us to reformulate distributional regression as a more general and flexible DistrNN. The functional form of this new network is driven by the input data, and we can relax the existing assumptions about the data distribution, the parametric model, and the stationarity of the data. The proposed DistrNN, as a feedforward network, is a hierarchical chain of layers representing high-dimensional and/or nonlinear input variables with the aim of predicting the target output variable. Importantly, we approximate the conditional distribution function with multiple network outputs as a set of joint probabilities. 

As a crucial step, instead of making assumptions about the function approximation probabilities, we define the model as an unknown general function that must be approximated by a neural network. Next, we consider a set of probabilities corresponding to $\Big\{\alpha_1,\ldots,\alpha_k\Big\}$, which are $k$ regularly spaced levels that characterize the conditional distribution function, and model them jointly as
\begin{equation}
\Big\{ \Pr\Big(p_{t,h}\le q_h^{\alpha_1}|\boldsymbol{z}_{t-1}\Big),\ldots,\Pr\Big(p_{t,h}\le q_h^{\alpha_k}|\boldsymbol{z}_{t-1}\Big) \Big\} = \mathfrak{g}_{W,b,h}(\boldsymbol{z}_{t-1}),
\label{eq:ffnet}
\end{equation}
where $\mathfrak{g}_{W,b,h}$ is a multiple-output neural network with $L$ hidden layers, which we name the DistrNN:
\begin{equation}
 \mathfrak{g}_{W,b,h}(\boldsymbol{z}_{t-1}) = g^{(L)}_{W^{(L)},b^{(L)}} \circ \ldots \circ g^{(1)}_{W^{(1)},b^{(1)}} \left(\boldsymbol{z}_{t-1}\right),
\label{eq:ffnetDistr}
\end{equation}
where $g^{(1)},\ldots,g^{(L)}$ is a collection of nonlinear activation functions with $W=\left(W^{(1)},\ldots,W^{(L)}\right)$ and $b=\left(b^{(1)},\ldots,b^{(L)}\right)$ as their weight matrices and bias vectors, respectively. Each weight matrix $W^{(\ell)} \in \mathbb{R}^{m\times n}$ contains $m$ neurons as $n$ column vectors $W^{(\ell)} = [w_{\cdot,1}^{(\ell)},\ldots,w_{\cdot,n}^{(\ell)}]$, and $b^{(\ell)}$ denotes the thresholds or activation levels. 

\begin{figure}[ht]
    \centering
    \includegraphics[width=\textwidth]{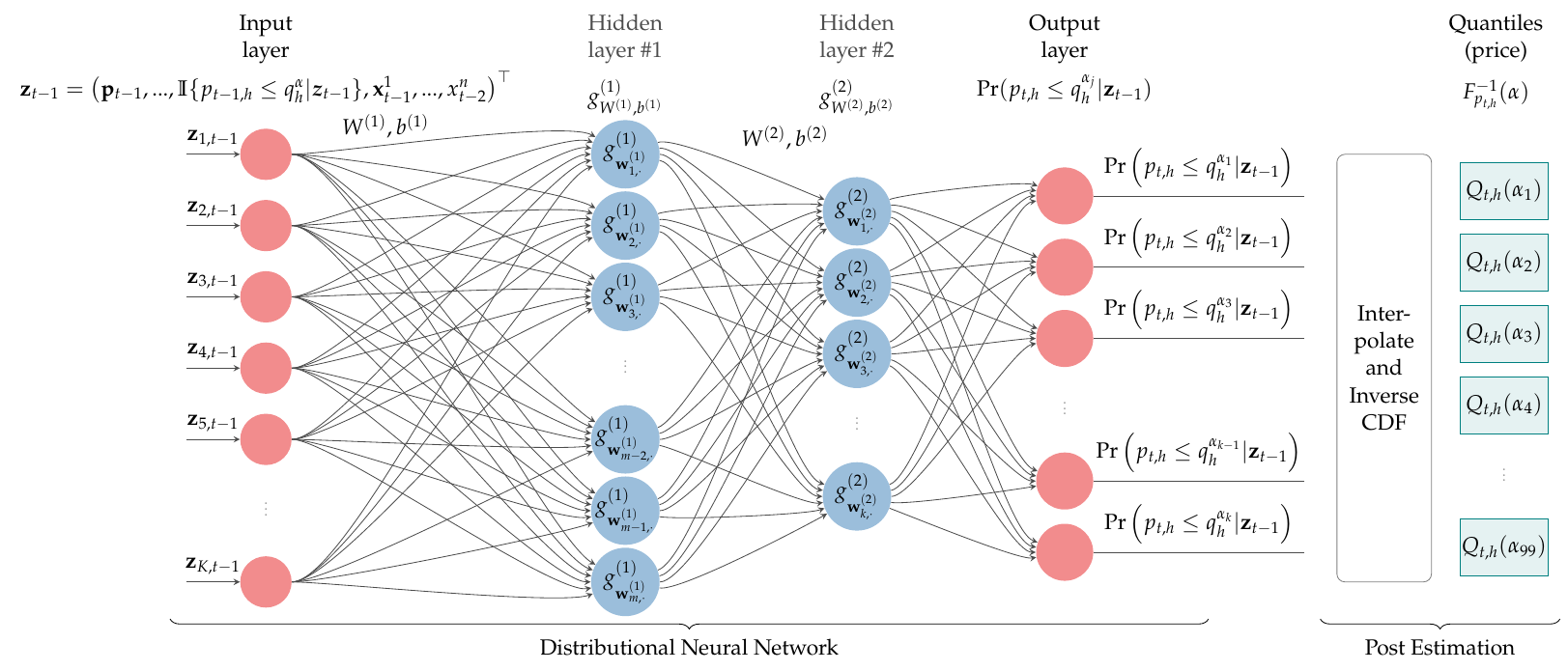}
    \caption{The DistrNN.} 
    \vspace{\medskipamount}
        \begin{minipage}{\textwidth} 
            \footnotesize
            An illustration of a DistrNN $\mathfrak{g}_{W,b,h}(\boldsymbol{z}_{t-1})$ for modelling a collection of conditional probabilities $\Big\{ \Pr\Big(p_{t,h}\le q_h^{\alpha_1}|\boldsymbol{z}_{t-1}\Big),\ldots,\Pr\Big(p_{t,h}\le q_h^{\alpha_p}|\boldsymbol{z}_{t-1}\Big) \Big\} $ with a set of predictor variables $\boldsymbol{z}_t $. The DistrNN is composed of two layers; however, it is conceivable that a substantial number of hidden layers may be incorporated into the network. The right-hand part of the figure elucidates the process used to derive the quantiles of day-ahead prices from the CDF provided by the network.
        \end{minipage}
    \label{fig:deepnet}
\end{figure}

Importantly, in contrast with the literature, we consider a multioutput (deep) neural network to characterize the collection of probabilities. Before discussing the estimation details that allow us to preserve the monotonicity of the probabilities, we illustrate the DistrNN in Figure~\ref{fig:deepnet}. We build a (deep) neural network $\mathfrak{g}_{W,b,h}(\boldsymbol{z}_{t-1})$ that is nonlinear, and the complexity of the network increases with the numbers of neurons $m$ and the number of hidden layers $L$. In its last layer, the DistrNN uses sigmoid activation to transform the outputs into probabilities. Note that for $L=1$, the neural network becomes a simple logistic regression model.

\subsection{Loss function}
\label{sec:loss}

Since we want to estimate the CDF, which is a nondecreasing function bounded within $[0,1]$, we need to design an objective function that minimizes the differences between the target distribution and the estimated distribution while imposing a nondecreasing property on the output. Given that this estimation task can be framed as a more complex classification problem that is analogous to logistic regression, we employ a binary cross-entropy loss function. In addition, we introduce a penalty term to enforce the ordering of the predicted probabilities in this multiple-output classification setting.

The loss function is then composed of two parts: the traditional binary cross-entropy loss and a penalty that adjusts for the monotonicity of the predicted output:
\begin{eqnarray}
\label{eq:loss-bince}
   \nonumber \mathcal{L} &=& - \frac{1}{T}\sum_t^{T} \underbrace{\frac{1}{k} \sum_j^k\left(\mathbb{I}\{p_{t,h}\le q_h^{\alpha_j}\}\log\left\{\mathfrak{g}_{W,b,h,j}(\boldsymbol{z}_{t-1})\right\} + \left(1-\mathbb{I}\{p_{t,h}\le q_h^{\alpha_j}\}\right)\log\left\{1-\mathfrak{g}_{W,b,h,j}(\boldsymbol{z}_{t-1})\right\} \right)}_\text{\text{binary cross-entropy}} \\
    &&+~ \underbrace{\lambda_{m} \sum_t^{T} \sum_{j=1}^{k-1}\left(\mathfrak{g}_{W,b,h,j}(\boldsymbol{z}_{t-1}) - \mathfrak{g}_{W,b,h,j+1}(\boldsymbol{z}_{t-1})\right)_+}_\text{\text{monotonicity penalty}}
\end{eqnarray}
where $(u)_+$ is a rectified linear unit (ReLU) function, i.e., $(u)_+ = \max\{u,0\}$, which passes through only positive differences between pairs of neighboring values ($j$ and $j+1$) of the CDF that violate the monotonicity condition, and $\mathbb{I}\{.\}$ is an indicator function. Such violations are controlled by the penalty parameter $\lambda_m$. Note that in addition to its simplicity, the ReLU function is employed for convenience, allowing for general use.\footnote{This choice enables the use of a GPU and hence increases the computational capacity that is available for solving more complex problems. The use of owned or nonoptimized functions for GPUs is not desired, and $(u)_+$ is commonly employed by libraries that work with GPUs.}

\section{Data}

Our analysis involves examining the hourly day-ahead electricity market in Germany, where day-ahead market prices play a critical role in the operations of day-ahead auctions. In this market, forecasts are utilized to formulate bids covering 24-hour periods. Specifically, on day $t-1$, market participants submit bids for each hour of the subsequent day. These bids are accepted until a specified deadline, after which the market undergoes a clearing process, and the participants receive energy allocations on the basis of the market clearing price. The utilized dataset spans from 1 January 2015 to 31 December 2023, and we obtained it from \citet{lipiecki2024postprocessing}.\footnote{The data are available at \url{https://transparency.entsoe.eu/} and \url{https://www.investing.com}.} Detailed descriptions of the dataset variables are provided in Sections~\ref{sub:input_variables}~and~\ref{sub:target_variable}.

\begin{figure}[ht]
    \includegraphics[width=.99\textwidth]{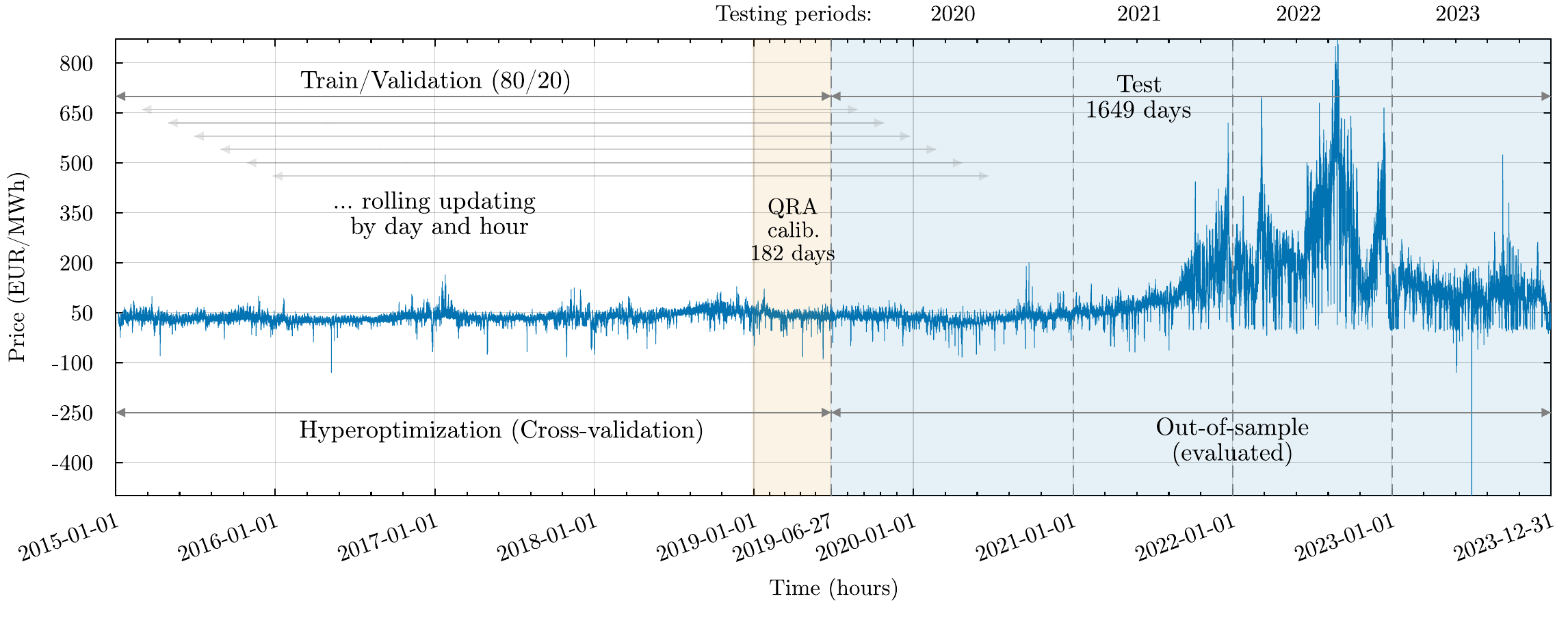}
    \caption{Electricity price data with distinguished and depicted periods for estimation purposes. The training and validation subsamples show how much of the data are used for training the model in both stages: hyperoptimization training and rolling window learning.}
    \label{fig:price-data-setup}
\end{figure}
To predict day-ahead electricity prices $p_{t,h}$ with hourly observations, we partition the data (following the standard practices) into training, validation and test sets. As illustrated in Figure~\ref{fig:price-data-setup}, the final 4.5 years serve as the out-of-sample (OOS) test set. The models are trained using the data preceding the OOS period, which are further split into training and validation sets. These subsets are employed for hyperparameter optimization or cross-validation to select parameters. Since auctions occur daily, all the models are estimated using a rolling window approach, shifting forward by one day (24 hours) at each step.

Additionally, Figure~\ref{fig:price-data-setup} highlights the quantile regression averaging (QRA) validation block, of which 182 days serve as a calibration window for the quantile regression model. This calibration window is excluded from the evaluation of the OOS results. The test period consists of 1649 days, spanning from June 27, 2019, to December 31, 2023. This period is further divided into four subperiods: the first, labeled ``2020'', also includes the second half of 2019,\footnote{This period covers June 27, 2019, to December 31, 2020, as in \citet{WeronMarcjasz:2023aa}.} while the remaining three subperiods correspond to the consecutive years of 2021, 2022, and 2023. This partitioning scheme aligns with the literature, particularly \citet{lipiecki2024postprocessing}.

% % __________ 
\subsection{Data transformation}

\begin{figure}[ht]
    \includegraphics[width=.99\textwidth]{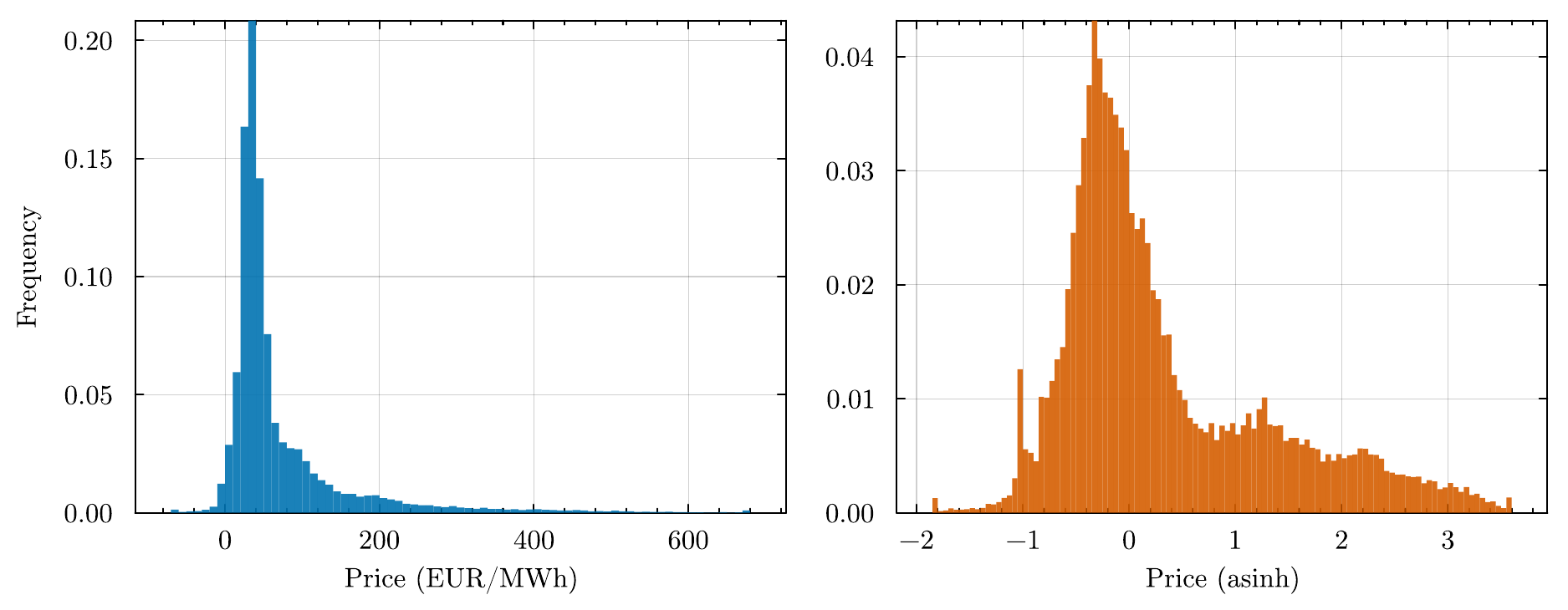}
    \caption{Empirical densities of the original and transformed price data (all hours).}
    \label{fig:price-hist-transformed}
\end{figure}

Prior to implementing the estimation procedure, we transform the data, as is common in the literature, to stabilize their variance and make the distribution more symmetrical. Since German electricity prices are allowed to be negative, we cannot use a logarithmic transformation. We adopt the variance-stabilizing transformation of \citet{Uniejewski:2018aa,narajewski2020econometric}, which is commonly employed in the literature. Before estimating the model, we preprocess the electricity prices with the \textit{area hyperbolic sine} (asinh) variance-stabilizing transformation
\begin{equation}
\text{asinh}(u) = \log\left( u + \sqrt{u^2+1}\right) 
\end{equation} 
where $u$ is the price standardized by subtracting the in-sample median and dividing the result by the median absolute deviation; it is adjusted by the 75\% quantile of the standard normal distribution to ensure its asymptotic consistency with the standard deviation. To recover the price forecasts, we apply the inverse transformation, i.e., the hyperbolic sine \citep{narajewski2020econometric,marcjasz2020probabilistic}. Figure~\ref{fig:price-hist-transformed} shows the empirical density of the day-ahead prices before and after applying the transformation, showing the changes exhibited by the scale and shape.

\subsection{Input variables}
\label{sub:input_variables}

On the basis of the data provided by \cite{lipiecki2024postprocessing,WeronMarcjasz:2023aa}, a consistent approach is used to construct the inputs for all the models. The input features correspond to the day-ahead price data acquired at their respective times, resulting in 24-hour-ahead prices denoted by $\boldsymbol{p}_t = [p_{t,1}, \dots p_{t,24}]$. The inputs $\boldsymbol{z}_{t-1}$ form the information set $\mathcal{I}_{t-1}$, which includes historical price data and other exogenous variables. 

However, as neural networks estimate time-dependent variables with complex and nonlinear relationships, it is still necessary to provide lagged time series inputs because of the autocorrelation of the data and the presence of seasonal patterns (such as daily and weekly patterns). Therefore, we begin by including the previous \textit{day-ahead prices}, specifically $\boldsymbol{p}_{t-1}$, $\boldsymbol{p}_{t-2}$, $\boldsymbol{p}_{t-3}$, and $\boldsymbol{p}_{t-7}$, as lags. Next, it is important to include variables that capture the dynamics of the probabilities, that is, lags of the indicator functions $\mathbb{I}\{p_{t-1,h} \leq q_h^{\alpha_j} | \boldsymbol{z}_{t-1}\}$ for all $j=1,\ldots,k$. The \textit{total load} variable is important in this study because it is a target variable. We include all hours of the day-ahead forecasts for the previous two days, including $\boldsymbol{x}_{L,t}$, $\boldsymbol{x}_{L,t-1}$ and $\boldsymbol{x}_{L,t-7}$. The last variable to be included in the 24-hour period relates to a day-ahead renewable energy resource forecast. For this purposes, we include data for the day ahead and the previous day, i.e., $\boldsymbol{x}_{R,t}$ and $\boldsymbol{x}_{R,t-1}$, respectively. Other external variables to be included are the closing prices of EU allowances (${x}_{E,t-2}$) and the prices of fuels, particularly coal, gas and oil (${x}_{C,t-2}$, ${x}_{G,t-2}$ and ${x}_{O,t-2}$, respectively). As we are forecasting $(t)$ today, these costs reflect the most recent available data (two days ago) according to the standard practices of the day-ahead auction market. Finally, to account for the weekly pattern exhibited by the data, we include a vector of weekday dummies, ${x}^{W}_{t,weekday}$, for the specific day of the week.\footnote{For the LEAR models, we use dummy variables to capture the days.} The total number of columns contained in the input matrix is 252. We consider the inputs $
\boldsymbol{z}_{t-1} = \big[\boldsymbol{p}_{t-1}, \boldsymbol{p}_{t-1}, \boldsymbol{p}_{t-3}, \boldsymbol{p}_{t-7}, \mathbb{I}\{p_{t-1,h} \leq q_h^{\alpha_1}  | \boldsymbol{z}_{t-1}\},\ldots,\mathbb{I}\{p_{t-1,h} \leq q_h^{\alpha_k}  | \boldsymbol{z}_{t-1}\},\boldsymbol{x}_{L,t}, \boldsymbol{x}_{L,t-1}, \boldsymbol{x}_{L,t-7}, \boldsymbol{x}_{R,t},\boldsymbol{x}_{R,t-1},{x}_{E,t-2}, {x}_{C,t-2}, {x}_{G,t-2}, {x}_{O,t-2}, {x}_{W,t,weekday}\big]$ for all the models except the naive model

% __________ 
\subsection{Target variables and information set}
\label{sub:target_variable}

To forecast the probability that the day-ahead electricity price will be below certain quantile levels, we model it as a set of probabilities on the basis of the conditional price information set. The accuracy of the forecasting results largely depends on the use of well-defined empirical quantiles $q_h^{\alpha}$, which correspond to a set of probabilities. Owing to location, size and shape differences that affect hourly price distributions, the target variable is determined on an hourly basis. The predicted variable is a set of hourly indicators that are related to $k$ equidistant probability levels $\alpha_j = \{0.01, \dots, 0.99\}$, where $k=31$.\footnote{We also experiment with different numbers of probability levels $k$, and while the results do not change, we use $k=31$ as a sufficient approximation.} The target variable is as follows:
\begin{equation}
 \mathbb{I}\{p_{t,h} \leq q_h^{\alpha_j}  | \boldsymbol{z}_{t-1}\}, \text{ for } h=1,...,24, \forall \alpha_j.
\label{eq:def_target}
\end{equation}
The unconditional quantiles defined by hours allow us to assume that the distribution within the information set is hour-specific.

Finally, the data for the target and input variables are subjected to \textit{winsorization}, and we use a proportion of 0.1\% to address the extreme minimum and maximum values included within the information set. Eq.~\ref{eq:def_target} shows that the CDF approximation approach uses unconditional quantiles $q^{\alpha_j}$ as the indicator of the target variable. Performing winsorization with a small fraction has no effect because the lowest and highest $\alpha$ values are greater than 0.1\%. Addressing extreme outliers, such as spikes or negative prices, is beneficial for the postestimation inverse transformation process, where we follow \citet{fritsch1980interpolation}. Notably, within the training and validation sets, which comprise 1440 observations, winsorization effectively handles one or two extreme values that could produce uninformative biases in the forecasts.

\section{Estimation}
\label{sec:estimation}

The developed forecasting frameworks and our nonparametric DistrNN approach are presented in this section. Furthermore, we outline several benchmark models, including a naive model, two quantile regression averaging models based on a linear specification with autoregressive and exogenous variables, and a DDNN benchmark \citep{nowotarski2018recent,serafin2019averaging,WeronMarcjasz:2023aa}. These models serve as robust and well-established baselines in the probabilistic electricity price forecasting literature. Since bids in electricity auctions are posted once per day for all hours, we follow the standard practice in the literature \citep{Maciejowska:2016aa,Liu:2017aa} by forecasting the distributions of the day-ahead prices for each hour using the same information set for a day.

\subsection{DistrNN}

\begin{figure}[ht!]
	\centering
    \includegraphics[width=.95\textwidth]{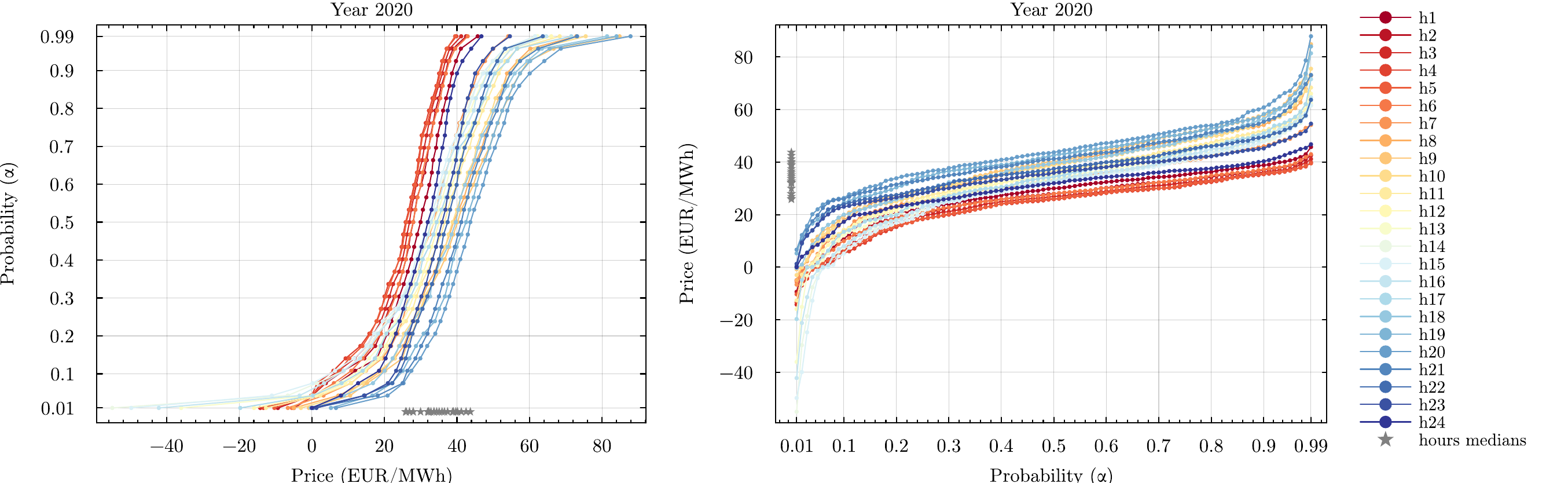}
    \includegraphics[width=.95\textwidth]{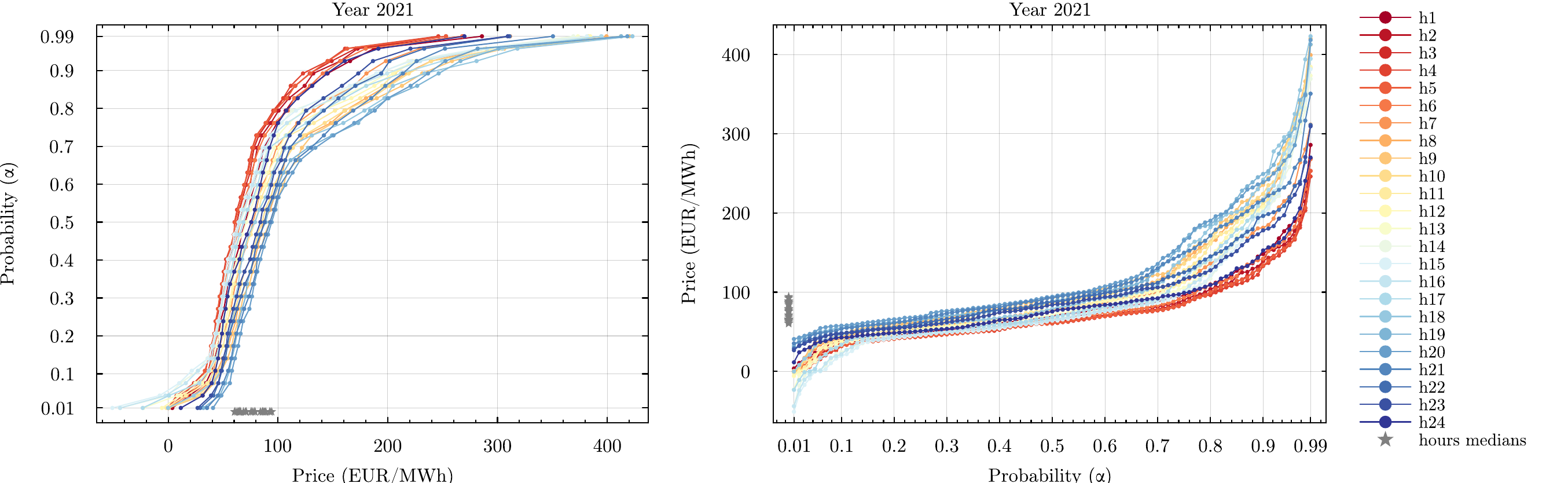}
    \includegraphics[width=.95\textwidth]{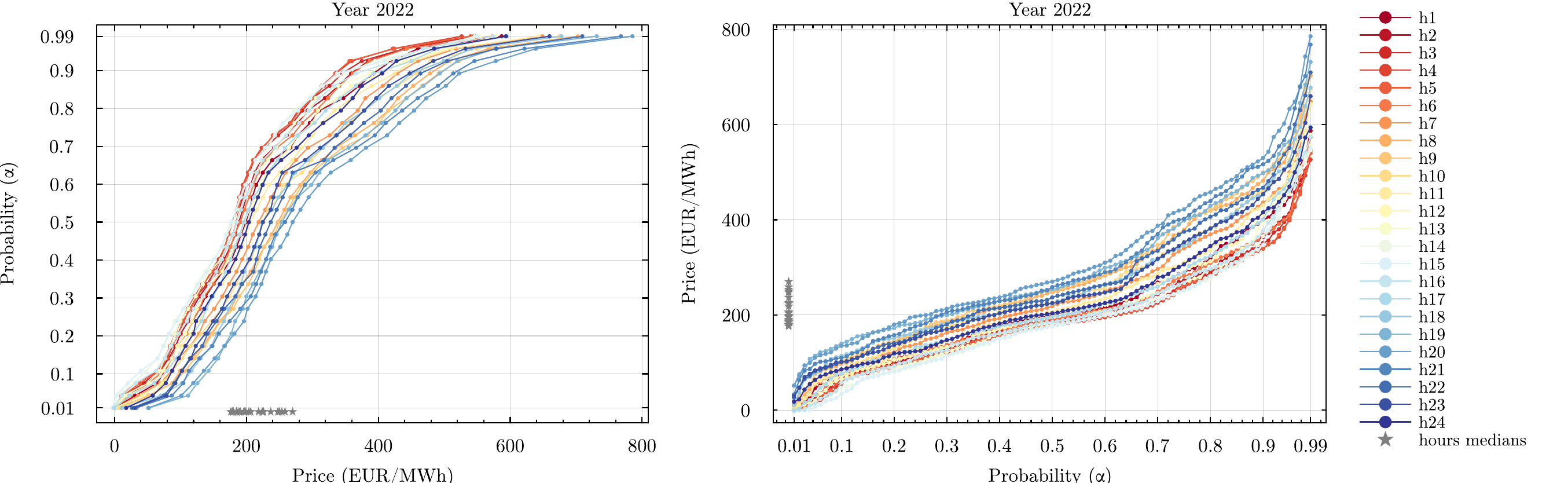}
    \includegraphics[width=.95\textwidth]{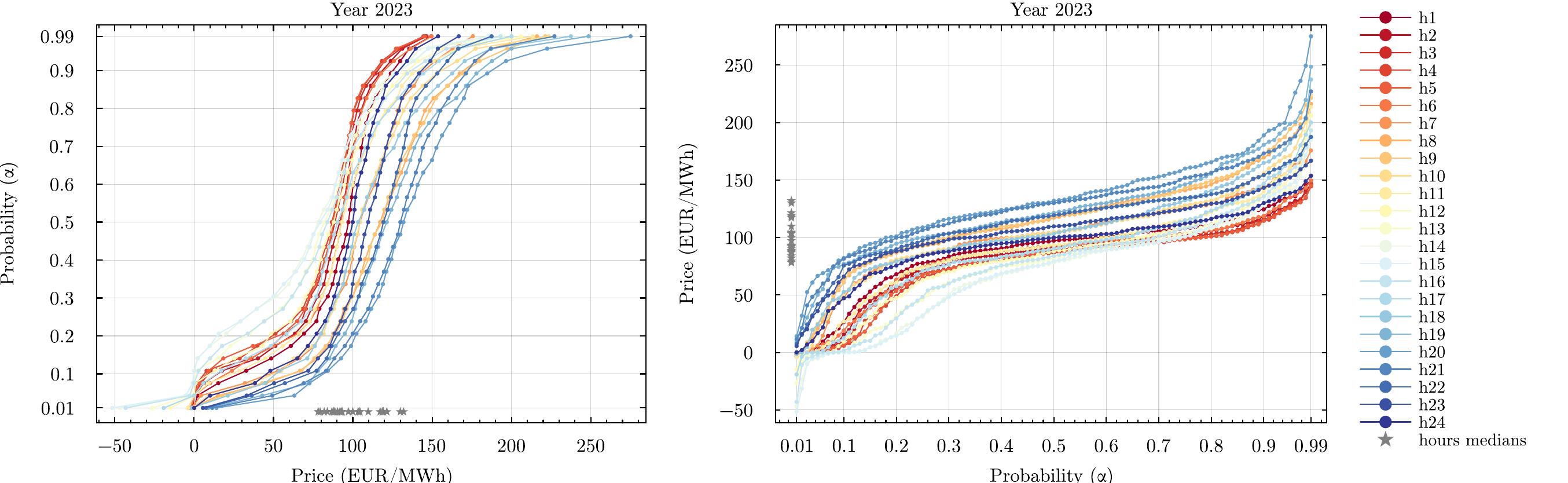}
    \caption{Unconditional functions of the cumulative distribution $F_h(p_{t,h})$ and quantile $Q_h(\alpha)$ by hours (represented by warm-to-cold colors) for each subperiod. Left: Illustration of the 31 $\alpha$ probability levels used to provide our target variable. Right: The unconditional quantiles produced for 99 $\alpha$ probability levels mimic the final distributional prediction results.}
    \label{fig:uncond-quantiles}
\end{figure}

Following the adage stating that \textit{``a picture is worth a thousand words''}, Figure~\ref{fig:uncond-quantiles} illustrates the relationship between the CDFs and the quantile functions, providing a visual representation of the distributional forecasting problem from left to right. The left panel shows the CDF approximations produced for each day and hour, which are constructed using $k$ equidistant points corresponding to $\alpha_j$ probabilities. The right panel presents the predicted 99 quantiles of the day-ahead prices for a given hour, which are obtained through inversion and interpolation. Both sides of the figures are based on the unconditional values of each subset of the dataset; however, they effectively demonstrate the structure of our results produced for a single day-ahead forecast and its associated 24 hours. We observe changes in both the shifts and shapes of the distribution for all hours.

Our DistrNN is implemented as a multilayer perceptron with two (possibly more) hidden layers, each containing a different number of neurons (chosen through a hyperparameter). The input size of the DistrNN is 252 features. We make no assumptions about the shapes of the distributions, as the DistrNN directly outputs the vector of probabilities, which includes 31 values that approximate the CDF. We allow for a different distribution for each hour, resulting in twenty-four DistrNNs to train.

To develop and analyze the model, we use the \textsc{Julia} programming language, specifically the Flux.jl package \citep{Flux2018}, for neural network training. Most neural networks have essential components such as optimization algorithms and techniques to prevent overfitting. We implement weighted adaptive moment estimation (AdamW) \citep{loshchilov2019adamw} as our optimization algorithm, which includes regularization techniques. AdamW mimics $L_2-norm$ regularization through its weight decay scheme as learning occurs. We then apply batch normalization to the weights of the first and second model layers and use an activation function and the dropout regularization method \citep{srivastava2014dropout}. The learning rate for stochastic gradient descent (Adam, \citet{kingma2014adam}), denoted as $\eta$; the weight decay parameter of regularization, denoted as $\lambda_W$; and the dropout proportion parameter, denoted as $\phi$, which indicates how many neurons should be switched off in each layer, are subjected to hyperoptimization.

\subsubsection{Training and hyperoptimized tuning}

Our forecasting procedure conforms to the standards of forecasting studies that use a forward rolling scheme on daily data. We utilize data acquired from a four-year (1440 observations) training and validation period (shown in Figure~\ref{fig:price-data-setup}) to perform a hyperparameter optimization search for determining the best model parameters. To reduce the incurred computational cost, hyperoptimization is performed before the rolling window scheme is implemented. We apply k-fold cross-validation on randomly shuffled data to increase the probability of model generalization rather than data storage. The dataset is divided into five cross-validation sets, with a 1:5 cross-validation ratio. 

For each hour of day-ahead prices, we determine the optimal DistrNN parameters. For hyperparameter optimization purposes\footnote{We use the Julia package \href{https://github.com/baggepinnen/Hyperopt.jl}{Hyperopt.jl}.} the algorithm considers 40 parameter combinations in a grid fashion on the basis of the parameter ranges and sets outlined in Table~\ref{tab:hyper-opt}. The best parameter set is selected based on the smallest average validation loss induced from cross-validation. 

The neural network is trained for a maximum of 1500 epochs prior to each year of the OOS period. We subsequently update the pretrained model for 500 additional epochs in the rolling window scheme rather than conducting full retraining for every hour of every day. Early stopping is applied if the validation loss exhibits no improvement. Additionally, the batch size chosen within the optimization scheme is either 32 or 64 data points. Furthermore, in terms of generalizability, the input data for training the DistrNN are augmented with $\mathcal{N}(0,0.1)$-distributed noise.
\begin{table}[ht]
  \centering
  {\smaller
  \begin{tabular}{lllllll}
    \toprule
    \textbf{Hyperparameters} & \textbf{Values} & & \textbf{Fixed parameters} & \textbf{Values} & \\
    \midrule
    Learning rate, $\eta$ & \multicolumn{2}{l}{$Range_1(0.0001, 0.003)$}& Number of HPO combinations & 40 & \\
    Dropout rate, $\phi$ & \multicolumn{2}{l}{$Range_2(0.0, 0.5)$}& Epochs & 1500 & \\
    $L_2$ decay rate, $\lambda_W$ & \multicolumn{2}{l}{$Range_2(0.000001, 0.01)$}& Early stopping patience & 100 & \\ % 50
    Number of hidden neurons in each layer & $Range_2(32,256)$ & & Monotonicity, $\lambda_m$ & 1.5 & \\
    Minibatch size & $\{32, 64\}$ & & Number of layers & 2 & \\
    Activation functions & \multicolumn{2}{l}{$\{$ReLU, tanh, sigmoid, } & $\alpha$ levels $(k)$ & 31 & \\
    \textbf{} & softmax, ELU$\}$ & & CV k-folds & 5 & \\
     & & & Ensembles & 4 & \\ \bottomrule
  \end{tabular}
  }
  \caption{The parameter values used to train the DistrNN.}%~\ref{}
    \begin{minipage}{\textwidth} 
      \footnotesize
      The hyperoptimization algorithm searches through the hyperparameter space and randomly tests a number of parameters sets to train the network. $Range_1$ is the evenly-spaced log range, and $Range_2$ is the evenly-spaced linear range.
    \end{minipage} 
  \label{tab:hyper-opt}
\end{table}

\subsubsection{Forward rolling procedure}

The objective of this study is to forecast day-ahead electricity prices with hourly granularity. To assess the performance of these forecasts, we utilize an OOS period comprising 1649 days, as shown in Figure~\ref{fig:price-data-setup}. The OOS period spans from June 27, 2019 to December 31, 2023. To generate forecasts, we adopt a forward rolling scheme that obtains hourly predictions for each day. This approach is consistent across all the models presented in this text, with any model-specific nuances addressed in the relevant sections.

Next, we outline the complete scheme of the DistrNN model and the process of deriving distributional forecasts for day-ahead prices. Initially, four years of historical data, covering the training and validation periods prior to the first day of the OOS period, are utilized to separately identify the optimal parameter set for each hour. This is achieved through hyperparameter optimization, which is implemented via a grid search. As a criterion, we minimize the binary cross-entropy function (Eq.~\ref{eq:loss-bince}). Given that the OOS period is split into four reporting subperiods and to optimize the computational efficiency of our method, the model is fully trained four times (once prior to each subperiod). Following this step, the model is updated daily within each subperiod using the tuned parameter set for each hour by employing a rolling window approach. To enhance the generalizability of the model, the four years of data used for training and validation are randomly shuffled and subsequently split into training and validation sets at an 80\%/20\% ratio. To mitigate the forecasting variance, the model is initialized multiple times, generating an ensemble of predictions. The final forecast is obtained by taking the top half of the ensemble results on the basis of the induced validation loss. Specifically, if the ensemble size is denoted as $n$, we select the first half, i.e., $\{\widehat{F}_1, \widehat{F}_2, \dots, \widehat{F}_{n/2}\}$, in this study, corresponding to two out of four predictions.

As shown in Figure~\ref{fig:deepnet}, the DistrNN predicts a CDF $\widehat{F}_{t,h}(\cdot|\boldsymbol{z}_{t-1})$, which is employed to find its inverse $\widehat{F}^{-1}(p_{t,h}) = \widehat{Q}_{t,h}(\alpha)$, namely, the quantile function, which is a collection of quantiles for a given $\alpha_j = \{0.01, \dots, 0.99 \}$. Before the inversion process, we use the monotonic cubic interpolation scheme from \citep{fritsch1980interpolation}. For the interval [0, 1], we obtain a monotonically increasing CDF via a finite grid consisting of 400 points. The inverse function is then found before aggregating the CDFs predicted by the DistrNN, which are denoted by $\widehat{F}^i_{t,h}$. In the context of neural networks, the aggregated ensemble mean is equivalent to the mean of the predictions. The predictions are averaged over the quantiles, which is defined as $\bar{Q}(\alpha) = \frac{1}{N_{ens}} \sum_i^{N_{ens}} \widehat{F}^{-1}_i(p)$, and $\bar{Q}_{t,h}(\alpha)$ is evaluated as our result. Notably, ensembles calculated over quantiles are considered instead of probabilities, although both quantile and probability averaging based on OOS ensembles yielded similar results in \citet{WeronMarcjasz:2023aa}. 

The complete procedure is repeated four times (runs), allowing for a comparative assessment of the DistrNN against the existing neural network-based electricity price forecasting approaches. This comparison highlights the benefits of employing model predictions averaged across multiple steps, which improves both the predictive accuracy and distribution stability of the model \citep{lago2021forecasting,Marcjasz2020,Weron:2014aa}. We report the results produced by the tested neural network models in two situations: the best results out of four runs (DistrNN$_\text{run}$) and the average of these runs (DistrNN$_\text{avg}$).

As detailed in Section~\ref{sec:loss}, the estimation process poses the inverse problem of quantile crossing: a possible violation of the monotonicity of the CDF. To solve this problem, we propose a loss function (Eq.~\ref{eq:loss-bince}) that explicitly penalizes such violations.

\subsection{Naive benchmark}

To remain consistent with the literature, we use the naive model in this paper. This model, as its name suggests, is a simple approach for predicting the price distribution on the next day using the prices of the previous day or week. Once the price point forecasts are available, one can bootstrap the price distribution from the errors between the predicted price and the true price for a given day \citep{nowotarski2018recent,Ziel:2018aa,Uniejewski:2018ab,WeronMarcjasz:2023aa}. The expected price for day $t$ and hour $h$ is
\begin{singlespace}
\begin{equation}
\mathbb{E}\left({p}_{t,h}\right) = \begin{cases}{p}_{t-7,h} & \text {for Monday, Saturday, and Sunday, }\\
{p}_{t-1,h} & \text {for Tuesday, Wednesday, Thursday, and Friday.}
\end{cases}
\label{eq:naive-price}
\end{equation}
\end{singlespace}
Then, the errors for one day, i.e., $\widehat{\varepsilon}_{t,h} = p_{t,h} - \widehat{p}_{t,h}$, are bootstrapped and added to the prices of Eq.~\ref{eq:naive-price} such that
\begin{equation}
\widehat{p}_{t,h}^i = \mathbb{E}(p_{t,h}) + \widehat{\varepsilon}_{t,h}^i, \text{ for } i \in 1,...,M.
\end{equation}
This provides naive distributional forecasts from the sampled prices of the model labeled Naive-B. Furthermore, we use another naive benchmark that assumes Gaussian innovations. Thus, the probabilistic forecasts are drawn from $N(0,\hat{\sigma})$, where $\hat\sigma$ is the standard deviation of $\widehat{\varepsilon}_{t,h}$ in a particular calibration window. This model is labeled Naive-1N, and we consider only one calibration window spanning 182 days.\footnote{For Naive-1N, we follow the implementation using the code of \citet{lipiecki2024postprocessing}.}

\subsection{The LEAR-QRA and LEAR-QRM benchmarks}

Given the objective of distributional electricity price forecasting, we consider two widely recognized approaches that are popular in the literature: QRA, as introduced by \citet{Nowotarski:2015aa}, and the quantile regression committee machine (QRM) of \citet{marcjasz2020probabilistic}. Both of these quantile regression models require point forecasts as inputs for estimating probabilistic forecasts. Among the models considered for point forecasting, the least absolute shrinkage and selection operator (LASSO)-estimated autoregressive (LEAR) model introduced by \citet{Uniejewski:2016aa} has been identified as one of the most accurate linear approaches \citep{Lago:2021aa}. Consequently, this state-of-the-art model is adopted as a sufficient parametric benchmark \citep{Mpfumali:2019aa,Uniejewski:2019aa,Zhang:2018aa,WeronMarcjasz:2023aa,lipiecki2024postprocessing}.

The point forecasting-based LEAR model, as detailed by \citet{Lago:2021aa}, is a linear regression model that uses a range of autoregressive and exogenous variables, e.g., section input variables, with parameters estimated through LASSO regularization \citep{tibshirani1996regression}. To maintain consistency with the literature and ensure reproducibility, the model specifications adhere to those outlined in \citep{WeronMarcjasz:2023aa,lago2021forecasting}. Specifically, the LEAR model is implemented using a forward rolling window scheme, where the window lengths for the input information set are 56, 84, 1092 and 1456 days \citep{Nowotarski:2015aa}. Parameter estimation is conducted via a 7-fold cross-validation scheme, which is aimed at identifying the optimal shrinkage parameter $\lambda$ across a grid of 100 candidate values using the least-angle regression algorithm \citep{efron2004least}.\footnote{Alternative model criteria, such as the Akaike information criterion or the Bayesian information criterion, could also be considered.} Given the defined window sizes, four distinct sets of point forecasts are generated, each covering a period of 1649+182 days. These point forecasts are subsequently used in the QRA and QRM rolling estimation schemes to derive probabilistic forecasts. Model estimation is performed independently for each hour $h$, while the information set remains consistent for each day $t$.

Both the QRA method and the QRM are estimated via quantile regression frameworks \citep{koenker1978regression}. This strategy provides a prediction of the conditional $\alpha$-quantile of $p_{t,h}$, $Q_{t,h}(\alpha)$, based on a set of regressors. In the case of QRA, the regressors include an intercept term and four point forecasts acquired from the LEAR model, which correspond to different window lengths: $\left[1,~\widehat{p}_{t,h}^{\{56\}},~\widehat{p}_{t,h}^{\{84\}},~\widehat{p}_{t,h}^{\{1092\}},~\widehat{p}_{t,h}^{\{1456\}}\right]$.
These configurations enable quantile averaging to be performed across multiple calibration windows, forming a method termed LEAR-QRA. 
Conversely, the QRM approach involves computing the average of the LEAR point forecasts, i.e., $\bar{p}_{t,h} = \frac{1}{4} (\widehat{p}_{t,h}^{\{56\}} + \widehat{p}_{t,h}^{\{84\}} + \widehat{p}_{t,h}^{\{1092\}} + \widehat{p}_{t,h}^{\{1456\}})$, which serves as the primary input for the quantile regression process; thus, this method is termed LEAR-QRM. The estimation processes of both methods are conducted by minimizing the quantile loss function (Eq.~\ref{eq:quantile-loss}) for each $\alpha$ quantile. To approximate the conditional distribution of the day-ahead electricity price, we estimate 99 quantiles. Through a forward rolling scheme, we use a window length of 6 months (182 days) to obtain 1649 days of OOS forecasts. While \citet{serafin2019averaging} suggested that the QRM generally outperforms QRA, this superiority is not universal and is not necessarily true for every evaluation metric, as shown by \citet{WeronMarcjasz:2023aa}. We estimate and report the results of LEAR-QRM in accordance with \citet{lipiecki2024postprocessing}.\footnote{We thank the authors of \citet{LIPIECKI2025102200} for providing the code for PostForecasts.jl~(\url{https://github.com/lipiecki/PostForecasts.jl}).}

\subsection{DDNN benchmark}

In addition to quantile regression approaches, we consider the DDNN introduced by \citet{WeronMarcjasz:2023aa} (DDNN-JSU), which represents the current state-of-the-art approach for conducting distributional electricity price forecasting using neural networks. This model was designed to learn the four parameters of Johnson's SU distribution for each hour, enabling the generation of a full probabilistic distribution for day-ahead electricity prices.\footnote{The replication code and hyperparameter optimization sets are available at \url{https://github.com/gmarcjasz/distributionalnn}.}
Consistent with the study of \citet{lipiecki2024postprocessing}, we adopt the hyperparameters selected by \citet{WeronMarcjasz:2023aa} for the period ending on December 31, 2018. This decision is motivated by two factors: the computational cost (weeks) already invested in hyperparameter optimization by the authors and empirical evidence suggesting that it is possible to transfer parameters across different electricity markets\citep{Marcjasz2020,lipiecki2024postprocessing}. In this study, the OOS period for the German electricity market is extended by three years beyond the dataset used to determine the optimal parameter set, thereby ensuring its applicability. Consistent with the original methodology, we perform a single run for each of the hyperparameter-optimized models to reproduce the results of DDNN-JSU as closely as possible. In Section~\ref{sec:results}, we present the best single run among the four executed models (DDNN-JSU$_\text{run}$), as well as the average of these predictions (DDNN-JSU$_\text{avg}$).

To enhance the complementarity of the DistrNNs to achieve a mutual benefit, we provide an average of the individual runs exhibited by both the DistrNN and DDNN-JSU, which is labeled as NN$_\text{avg}$.

\subsection{Evaluation criteria}

We evaluate the quality of the output probabilistic forecasts by focusing on the sharpness of the distributions yielded by all the models (following \citet{gneiting2007strictly}) via the continuous rank probability score (CRPS) measure:
\begin{equation}
CRPS_{t,h}(\widehat{F}_{p_{t,h}},p_{t,h}) = \int_{\mathbb{R}} \left(\widehat{F}_{p_{t,h}}(z) - \mathbb{I}\{p_{t,h} \leq z\}\right)^2 dz,
\end{equation}
where $\mathbb{I}\{p_{t,h} \leq z\}$ is the indicator function. As is common in the literature, we use the discrete approximation of the CRPS as shown below:
\begin{equation}
CRPS_{t,h} = \frac{1}{N_{\alpha}} \sum_{\alpha = 0.01}^{0.99} QL_\alpha(\widehat{q}^{\alpha}_{t,h},p_{t,h}),
\end{equation}
where $QL_{\alpha}$ is the $\alpha$-quantile loss function (or pinball loss), which we state as
\begin{equation}
QL_{\alpha}(\widehat{q}^{\alpha}_{t,h},p_{t,h}) = \left(\mathbb{I}\{p_{t,h} \leq \widehat{q}^{\alpha}_{t,h}\} - \alpha\right)\left(\widehat{q}^{\alpha}_{t,h} - p_{t,h}\right),
\label{eq:quantile-loss}
\end{equation}
where $\alpha$ is the probability, $\widehat{q}^{\alpha}_{t,h}$ is the quantile prediction obtained from $\widehat{F}_{t,h}$, $p_{t,h}$ is the original time series, and $N_{\alpha}$ is the number of quantile probability levels for which we approximate the quantile function from the CDF. In this way, we approximate the CRPS sum of the pinball scores produced over the discrete set of $\alpha = \{0.01, 0.02, ..., 0.99\}$ for all OOS values.

To assess the significance of the forecasting accuracy and the performance differences among the tested models, we use the Diebold--Mariano (DM) test \citep{diebold1995comparing} (the version with an adjusted Newey--West variance). With the DM test, we apply two testing approaches. First, we compare the errors induced by the models on a day-ahead basis, and second, we test the disaggregated accuracy achieved for each of the 24 hours. In our evaluation, the loss calculated for model $m$ and hour $h$ is denoted as $L_{m}^{t,h}$, i.e., the vector of the CRPS loss. In the overall test, we aggregate the losses to days where $L_m^t = \sum_{h=1}^{24} L_m^{t,h}$ and measure the statistical significance of the differences between all model pairs. The null hypothesis concerning two models is that the difference between the $L_1-norm$ values of the models is less than or equal to zero. This is denoted as $\mathcal{H}_0: \mathbb{E} [\Delta_{m_1, m_2}^t ] \leq 0$, where formally, $\Delta_{m_1, m_2}^t = ||L_{m_1}^t||_1 - ||L_{m_2}^t||_1$. Let us consider the disaggregated differences between the accuracies of the models so that for $h=1,...,24$ with losses of $L_{m}^{t,h}$, we test the null hypothesis $\mathcal{H}^h_0: \mathbb{E} [\Delta_{m_1, m_2}^{t,h} ] \leq 0$, where $\Delta_{m_1, m_2}^{t,h} = ||L_{m_1}^{t,h}||_1 - ||L_{m_2}^{t,h}||$. The null hypotheses of both tests are examined against the alternative stating that $m_2$ is more accurate than $m_1$ \citep{clements2008quantile,nowotarski2018recent}.\footnote{To perform the DM test, we use \url{https://github.com/JuliaStats/HypothesisTests.jl}.}

% % __________ 
\section{Results}
\label{sec:results}

We present empirical results for all the forecasting models applied to German day-ahead electricity prices, including both established benchmark models and our newly proposed model. The performance of each model is assessed using the CRPS complemented by statistical significance tests (e.g., the DM test) to implement forecasting accuracy comparisons. Following the common practices employed in the literature, we conduct the comparisons over four distinct subperiods: 2020, 2021, 2022, and 2023. This temporal segmentation scheme enables us to assess how each model performs under various market conditions and to highlight the potential structural changes exhibited by price dynamics over time.

We further perform a granular analysis of the achieved forecasting accuracy by examining the CRPS values and DM test results, disaggregated by hour of the day and by subperiod. This level of detail reveals how the prediction accuracies vary across different times of day and across different years, offering insights into the specific strengths and limitations of each modeling approach. Additionally, for the neural network models, we report both their single-run results and their average performance across multiple runs. As established in the forecasting literature, averaging predictions over multiple runs (ensemble averaging) improves both the forecasting accuracy and the distributional stability of the predictions \citep{lago2021forecasting, Marcjasz2020, Weron:2014aa}. This practice helps mitigate the randomness that is inherent in neural network training process and yields more reliable probabilistic forecasts.

\begin{figure}[ht!]
  \centering
    \includegraphics[width=.99\textwidth]{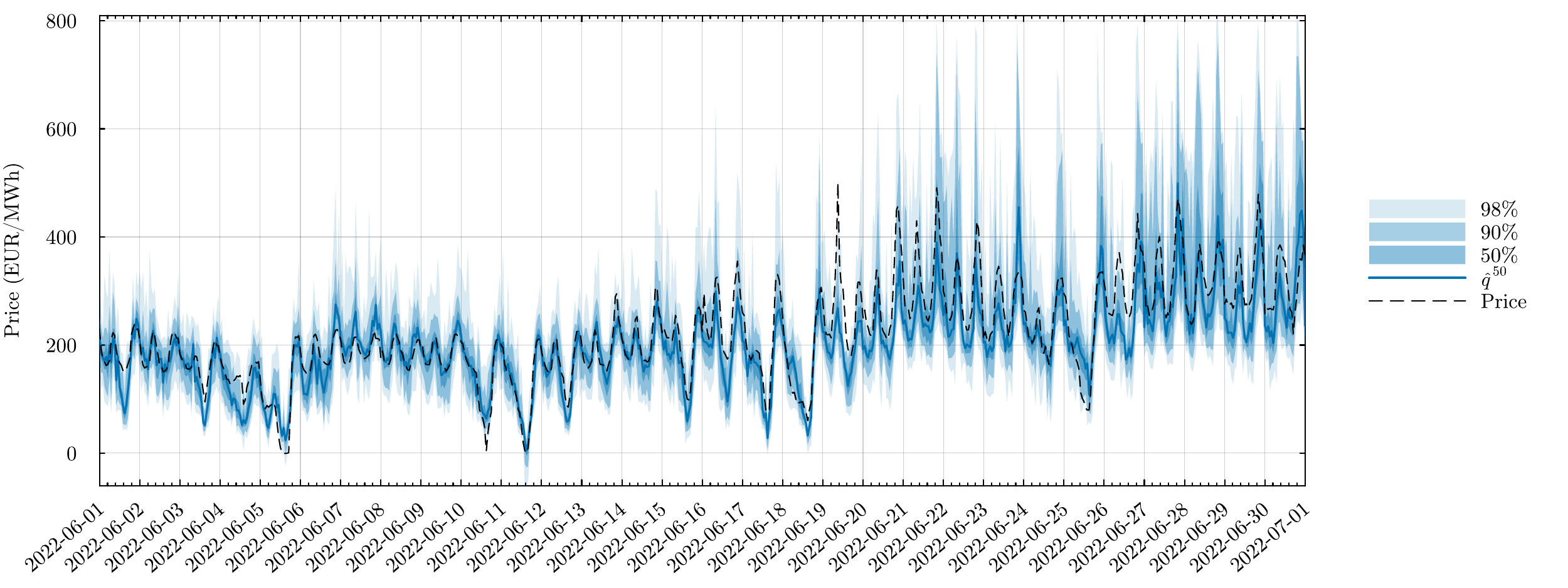}
  \caption{Examples of probabilistic day-ahead electricity price forecasts provided by the DistrNN. The figure shows the hourly forecasts produced for June 2022.}
  \label{fig:example-prediction}
\end{figure}

To illustrate the nature of the predictions, Figure \ref{fig:example-prediction} presents an example of the probabilistic electricity price predictions obtained for a selected period. The figure shows that the proposed model produces asymmetric probability forecasts that respond to the variations exhibited by the data and shifts in the underlying distribution. In the illustrated period, the prices exhibit a clear upward trend, introducing nonstationary behavior that is inherently more challenging to forecast. Consistent with this challenge, we observe that during this rising-price episode, all the models tend to slightly underestimate the actual prices. This underestimation might be attributed to the use of a rolling forecasting scheme based on unconditional in-sample quantiles, which can lag in terms of adjusting to a rapidly increasing price trend.

Furthermore, the results of the hyperparameter optimization process are presented in the appendix in Tables~\ref{tab:node_activations} and~\ref{tab:params_hpo}. The selected hyperparameter values are specific to each hour since the DistrNN is trained separately for each hour; hence, the range of optimal sets is wide. The optimal set obtained for each hour is then used in the forward rolling scheme to obtain OOS distributional predictions.

\subsection{OOS evaluation}

Table~\ref{tab:all-crps} presents an overview of the results produced by all the models, showing the CRPS values yielded by different probabilistic forecasting models from 2020\footnote{Note that this is a 1.5-year period from 27 June 2019 to 31 December 2020, which includes one-third of the data.} to 2023, with lower values indicating better probabilistic accuracy. Initially, in the case of deep neural network models, we report two cases for each model. First, we provide the results obtained for the single best run of the model, $run$, and second, we provide the ensemble average of 4 runs, $avg$, which is a standard metric and provides more stable results. Furthermore, to examine the averaging process used by the NN models, we report the average of one initialization of the two models as NN$_\text{avg}$, which is the average of DDNN-JSU$_\text{run}$ and DistrNN$_\text{run}$. This combination shows that each model provides different information about the underlying distribution.

Starting with the benchmark models, Naive-B and Naive-1N consistently exhibit the highest CRPS values across all subperiods. This indicates their limited capacity to adapt and provide good accuracy in situations with distributional dynamics that change over time.

In contrast, among the more informed models, LEAR-QRM and LEAR-QRA achieve good accuracies, especially in years with strong shifts exhibited by the underlying distributions. LEAR-QRM performs well in terms of its CRPS values for 2021 (4.189), 2022 (10.651), and 2023 (4.422). These results are likely attributed to the following factors. First, a more efficient algorithm is used in LEAR-QRM \citep{LIPIECKI2025102200},\footnote{The LEAR-QRA model is estimated following \citet{WeronMarcjasz:2023aa}, whereas LEAR-QRM uses a new package from \citet{LIPIECKI2025102200}.} and second, importantly, the electricity price within these years was affected by the war in Ukraine, which caused a trend and unstructured turbulence that the linear model with quantile regression captures well \citep{lipiecki2024postprocessing}. 
\begin{table}[htbp]
  \caption{Overall probabilistic forecasting performance measured by the CRPS metric across the years 2020--2023, evaluated over all 99 quantiles. The table compares the tested methods, with cell shading indicating their relative performance (green = lower CRPSs, red = higher CRPSs). Subscripts \textit{run} indicate the results derived from a single representative run, whereas \textit{avg} reports the average performance achieved across multiple runs, and NN$_\text{avg}$=(DDNN-JSU$_\text{run}$+DistrNN$_\text{run}$)/2. \textit{Note:} Models with $^{\star}$ follow the implementation of \citet{WeronMarcjasz:2023aa}, while models with $^{\diamond}$ follow \citet{lipiecki2024postprocessing}. The period reported as ``2020'' specifically captures the data between June 27, 2019, and December 31, 2020.}
	 \centering
	 \begin{tabular}{lcccc}
	\toprule
	\textbf{Model} & \textbf{2020} & \textbf{2021} & \textbf{2022} & \textbf{2023} \\
	\midrule
	Naive-B & \cellcolor[rgb]{ .973,  .412,  .42} 3.647 & \cellcolor[rgb]{ .973,  .412,  .42} 10.457 & \cellcolor[rgb]{ .976,  .478,  .435} 23.869 & \cellcolor[rgb]{ .98,  .529,  .443} 10.569 \\
	Naive-1N$^{\diamond}$ & \cellcolor[rgb]{ .976,  .435,  .427} 3.550 & \cellcolor[rgb]{ .98,  .518,  .443} 9.494 & \cellcolor[rgb]{ .973,  .412,  .42} 25.346 & \cellcolor[rgb]{ .973,  .412,  .42} 12.078 \\
	LEAR-QRA$^{\star}$ & \cellcolor[rgb]{ 1,  .871,  .51} 1.649 & \cellcolor[rgb]{ .69,  .831,  .498} 4.933 & \cellcolor[rgb]{ .682,  .827,  .498} 12.035 & \cellcolor[rgb]{ 1,  .922,  .518} 5.359 \\
	LEAR-QRM$^{\diamond}$ & \cellcolor[rgb]{ .576,  .796,  .49} 1.352 & \cellcolor[rgb]{ .388,  .745,  .482} 4.189 & \cellcolor[rgb]{ .388,  .745,  .482} 10.651 & \cellcolor[rgb]{ .388,  .745,  .482} 4.422 \\
	DDNN-JSU$_\text{run}$$^{\star}$ & \cellcolor[rgb]{ 1,  .918,  .518} 1.443 & \cellcolor[rgb]{ 1,  .914,  .518} 5.795 & \cellcolor[rgb]{ .996,  .843,  .502} 15.420 & \cellcolor[rgb]{ 1,  .878,  .51} 5.936 \\
	DistrNN$_\text{run}$ & \cellcolor[rgb]{ 1,  .922,  .518} 1.424 & \cellcolor[rgb]{ 1,  .914,  .518} 5.800 & \cellcolor[rgb]{ 1,  .886,  .514} 14.385 & \cellcolor[rgb]{ 1,  .91,  .518} 5.563 \\
	DDNN-JSU$_\text{avg}$$^{\star}$ & \cellcolor[rgb]{ .439,  .757,  .482} 1.329 & \cellcolor[rgb]{ 1,  .922,  .518} 5.691 & \cellcolor[rgb]{ 1,  .922,  .518} 13.521 & \cellcolor[rgb]{ .933,  .902,  .514} 5.260 \\
	DistrNN$_\text{avg}$ & \cellcolor[rgb]{ .71,  .835,  .498} 1.374 & \cellcolor[rgb]{ .941,  .902,  .514} 5.551 & \cellcolor[rgb]{ .988,  .918,  .514} 13.475 & \cellcolor[rgb]{ .922,  .898,  .51} 5.240 \\
	NN$_\text{avg}$ & \cellcolor[rgb]{ .388,  .745,  .482} 1.320 & \cellcolor[rgb]{ .627,  .812,  .494} 4.776 & \cellcolor[rgb]{ .91,  .894,  .51} 13.100 & \cellcolor[rgb]{ .894,  .89,  .51} 5.200 \\
	\bottomrule
	\end{tabular}%
  \label{tab:all-crps}%
\end{table}%
Among the neural network-based models, the ensemble neural model NN$_\text{avg}$ consistently ranks among the top-performing methods for every year. It achieves the best CRPS in 2020 (1.320) and remains competitive in the subsequent years, with a CRPS of 5.200 in 2023, second to that of LEAR-QRM. Given that NN$_\text{avg}$ is the average of DDNN-JSU$_\text{run}$ and DistrNN$_\text{run}$, we observe that this version generally performs worse than both NN$_\text{avg}$ and their averaged ensemble counterparts. Among the two models, our proposed DistrNN$_\text{run}$ model has lower CRPSs in the three subperiods, whereas DDNN-JSU$_\text{run}$ has a lower value of 5.795 in 2021. This finding remains true for the averaged ensemble variants, where the nonparametric DistrNN$_\text{avg}$ model achieves lower CRPSs in 2021, 2022, and 2023 than does the DDNN-JSU$_\text{avg}$ model. Both the DistrNN and DDNN-JSU averages outperform the LEAR-QRA model \citep{WeronMarcjasz:2023aa} on the 2020 and 2023 subsamples, but DDNN-JSU$_\text{avg}$ has a lower CRPS than does LEAR-QRM in the 2020 period.
During turbulent periods, linear quantile regression performs significantly better. These results highlight the advantages of ensemble methods in terms of reducing variance and improving predictive accuracy, even when solo runs of data-driven models are used.

\begin{figure}[ht!]
  \centering
    \includegraphics[width=.9\textwidth]{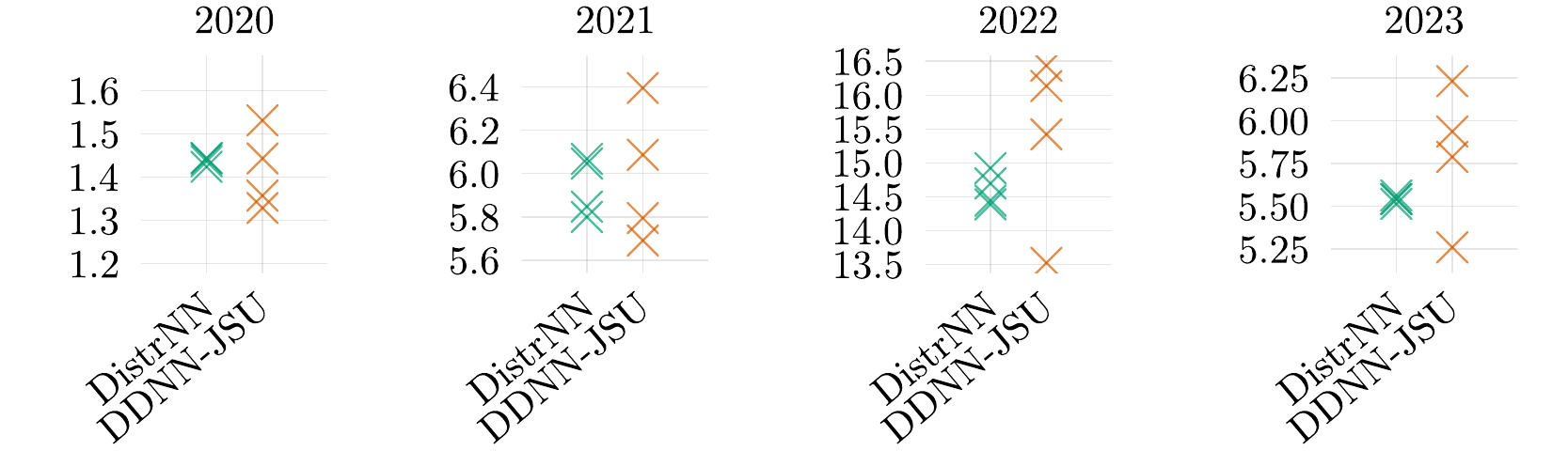}
  \caption{CRPS values derived from individual runs of the DistrNN and DDNN-JSU models across the four evaluation periods.}
  \label{fig:solo-crps}
\end{figure}
Figure~\ref{fig:solo-crps} compares the CRPS performances of four neural network models, with a particular focus on the performance of the DistrNN and DDNN-JSU over different periods. The DistrNN consistently achieves lower CRPS values, indicating stronger generalizability across different periods and market conditions. This performance may reflect lower within-sample variance across individual runs, contributing to greater predictive stability. However, the architectural differences between the models could also play a role. While DDNN-JSU supports wider networks with up to 1024 neurons per layer, the DistrNN is constrained to a maximum of 256 neurons. Despite this limitation, the lower CRPSs produced by the DistrNN suggest that it delivers comparable, if not superior, forecasting accuracy with a more compact network structure. This compactness has practical implications. In particular, the reduced memory usage of the DistrNN makes it advantageous for applications where computational resources are constrained, such as real-time or embedded forecasting environments. Notably, the DistrNN ensembles are generated from two (out of 4) predictions using selected parameter sets per run, which may reduce the variance observed within a single run. However, this does not fully account for the overall consistency of the model across different runs. Overall, the findings highlight the efficiency and robustness of the DistrNN in distributional forecasting tasks.

\begin{figure}[ht!]
  \centering
    \includegraphics[width=.59\textwidth]{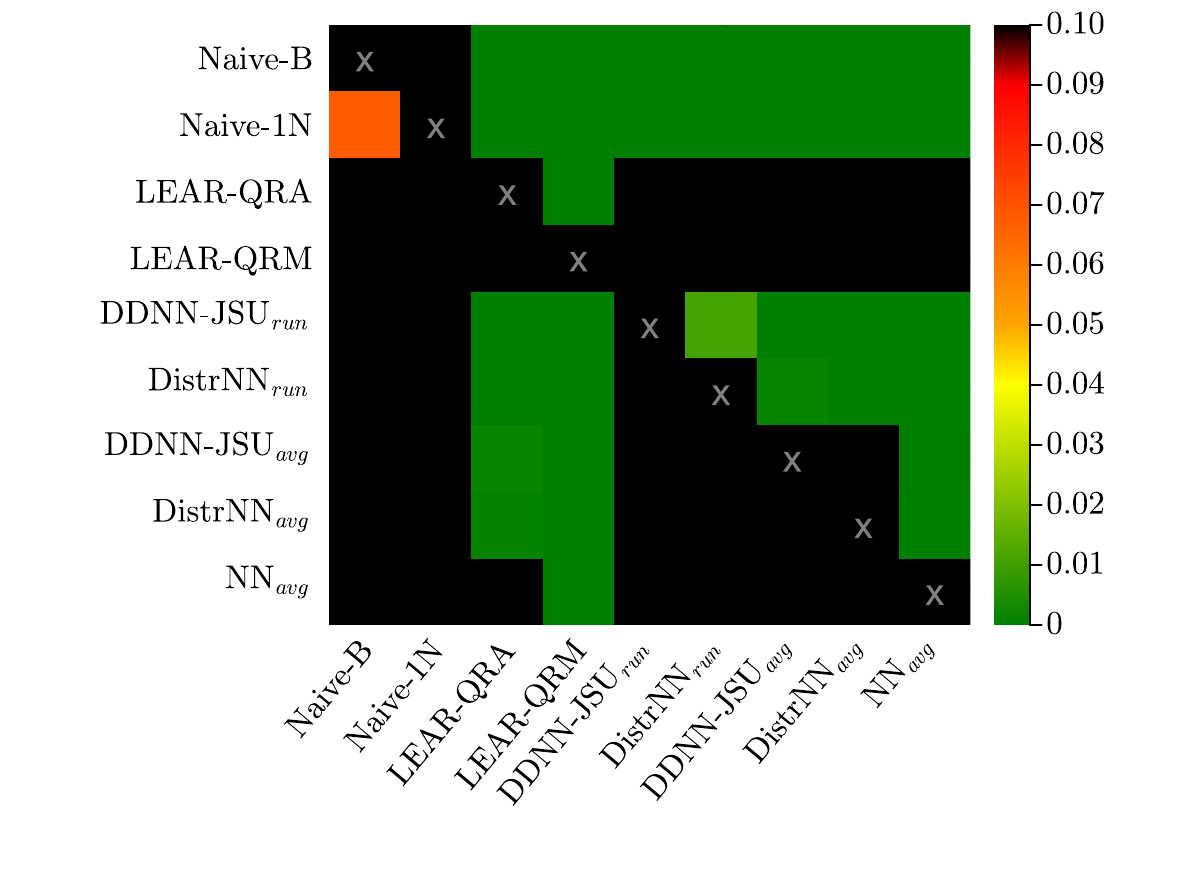}
  \caption{Pairwise DM test $p$-values for the CRPSs aggregated over all 99 quantiles for 2020--2023. The null hypothesis $\mathcal{H}_0$ states that the model on the \textit{y-axis} has a lower predictive loss than does the model on the \textit{x-axis}. Each cell shows the $p$-value produced for the corresponding model comparison. The colors of the cells (nonblack) indicate that the model on the \textit{x-axis} performs significantly better than the model on the \textit{y-axis} does at the 10\% significance level. The black cells indicate $p$-values that are greater than 10\%, implying that no statistically significant performance differences are observed.}
  \label{fig:dm-crps-matrix}
\end{figure}
The results of the DM test applied to the multivariate loss differences across the models are presented in Figure~\ref{fig:dm-crps-matrix}. The test evaluates pairwise predictive performances by comparing the absolute CRPS losses $L_m$ aggregated over 24 hourly forecasts for the OOS period.
In general, the DM test results presented in the table suggest that we reject the null hypothesis that any model is statistically better than LEAR-QRM, LEAR-QRA, and NN$_\text{avg}$, where the first result is better overall. 

Consistent with our expectations and Table~\ref{tab:all-crps}, all the models significantly outperform the naive benchmarks, as indicated by their significant $p$-values (green). This shows the importance of a structured model design for capturing the complexity of probabilistic electricity forecasting. Furthermore, DDNN-JSU${_\text{run}}$ and DistrNN${_\text{run}}$ exhibit improvements over the simpler benchmarks, where the DistrNN appears to be statistically better than DDNN-JSU. However, their averaged counterparts, DDNN-JSU$_\text{avg}$ and DistrNN$_\text{avg}$, yield stronger performance, as evidenced by the statistically significant differences favoring the ensembles. This highlights the benefits of averaging models to capture overall prediction distributions \citep{WeronMarcjasz:2023aa}. An interesting result of the averaging process is that the combination of the solo runs of the DistrNN and DDNN-JSU significantly outperforms the averaged prediction of each model, even when their individual averages are not significantly different. This result, along with the results acquired from Table~\ref{tab:all-crps}, provides interesting insight that both DistrNNs designed with different distributional objectives (parametric vs. nonparametric) can be leveraged in tandem to enhance the resulting performance, offering a principled approach for learning different parts of distributions, which offers a compelling case for hybrid ensembles.

\subsection{Performance across different hours and tails}

Table 3 reports the CRPS values achieved for the $1,\ldots,10$ and $90,\ldots,99$ quantiles, providing insight into the model performance in terms of capturing extreme tail events relative to their overall distributional accuracy. As expected, the naive models perform poorly across all years, confirming their limited utility in forecasting extreme outcomes. The neural network-based model averages (DDNN-JSU$_\text{avg}$, DistrNN$_\text{avg}$, and NN$_\text{avg}$) achieve the lowest CRPS scores in 2020, which is one-third of the OOS period. LEAR-QRM achieves the lowest CRPS values for the other two-thirds of the OOS years of 2021, 2022, and 2023. Interestingly, NN$_\text{avg}$ has second-best results in 2021 and 2023 (by $\approx$ 6~\% to 12~\%). This suggests that while traditional QRA methods remain effective under certain conditions, deep learning-based approaches offer valuable predictive capabilities for capturing extreme tail events.
\begin{table}[htbp]
  \centering
  \caption{Extreme-tailed probabilistic forecasting performance measured by the CRPS metric across the years 2020--2023, evaluated over the $1,\ldots,10$ and $90,\ldots,99$ quantiles. The table compares the tested methods, with the cell shading indicating relative performance (green = lower CRPSs; red = higher CRPSs). The \textit{run} subscripts indicate results derived from a single representative run, whereas \textit{avg} reports the average performance achieved across multiple runs, and NN$_\text{avg}$=(DDNN-JSU$_\text{run}$+DistrNN$_\text{run}$)/2. \textit{Note:} Models with $^{\star}$ follow the implementation of \citet{WeronMarcjasz:2023aa}, while models with $^{\diamond}$ follow \citet{lipiecki2024postprocessing}. The period reported as ``2020'' specifically captures the data between June 27, 2019, and December 31, 2020.}	\begin{tabular}{lcccc}
	\toprule
	\textbf{Model} & \textbf{2020} & \textbf{2021} & \textbf{2022} & \textbf{2023} \\
	\midrule
	Naive-B & \cellcolor[rgb]{ .973,  .412,  .42} 1.743 & \cellcolor[rgb]{ .973,  .412,  .42} 6.252 & \cellcolor[rgb]{ .973,  .412,  .42} 12.552 & \cellcolor[rgb]{ .98,  .518,  .443} 5.111 \\
	Naive-1N$^{\diamond}$ & \cellcolor[rgb]{ .976,  .42,  .424} 1.729 & \cellcolor[rgb]{ .984,  .604,  .459} 4.804 & \cellcolor[rgb]{ .98,  .51,  .439} 11.334 & \cellcolor[rgb]{ .973,  .412,  .42} 5.786 \\
	LEAR-QRA$^{\star}$ & \cellcolor[rgb]{ .996,  .835,  .502} 0.804 & \cellcolor[rgb]{ .788,  .859,  .502} 2.190 & \cellcolor[rgb]{ .745,  .847,  .502} 5.401 & \cellcolor[rgb]{ 1,  .914,  .518} 2.555 \\
	LEAR-QRM$^{\diamond}$ & \cellcolor[rgb]{ 1,  .922,  .518} 0.603 & \cellcolor[rgb]{ .388,  .745,  .482} 1.819 & \cellcolor[rgb]{ .388,  .745,  .482} 4.579 & \cellcolor[rgb]{ .388,  .745,  .482} 1.949 \\
	DDNN-JSU$_\text{run}$$^{\star}$ & \cellcolor[rgb]{ .788,  .859,  .502} 0.578 & \cellcolor[rgb]{ 1,  .922,  .518} 2.383 & \cellcolor[rgb]{ 1,  .886,  .514} 6.442 & \cellcolor[rgb]{ 1,  .922,  .518} 2.484 \\
	DistrNN$_\text{run}$ & \cellcolor[rgb]{ 1,  .922,  .518} 0.609 & \cellcolor[rgb]{ 1,  .902,  .514} 2.542 & \cellcolor[rgb]{ 1,  .898,  .514} 6.292 & \cellcolor[rgb]{ 1,  .914,  .518} 2.540 \\
	DDNN-JSU$_\text{avg}$$^{\star}$ & \cellcolor[rgb]{ .408,  .749,  .482} 0.535 & \cellcolor[rgb]{ 1,  .91,  .518} 2.495 & \cellcolor[rgb]{ 1,  .922,  .518} 5.979 & \cellcolor[rgb]{ .816,  .867,  .506} 2.326 \\
	DistrNN$_\text{avg}$ & \cellcolor[rgb]{ .776,  .855,  .502} 0.577 & \cellcolor[rgb]{ .953,  .906,  .514} 2.341 & \cellcolor[rgb]{ .961,  .91,  .514} 5.892 & \cellcolor[rgb]{ .835,  .875,  .506} 2.342 \\
	NN$_\text{avg}$ & \cellcolor[rgb]{ .388,  .745,  .482} 0.532 & \cellcolor[rgb]{ .529,  .784,  .49} 1.950 & \cellcolor[rgb]{ .808,  .867,  .506} 5.544 & \cellcolor[rgb]{ .678,  .827,  .498} 2.204 \\
	\bottomrule
	\end{tabular}%
  \label{tab:tails-crps}%
\end{table}%

\begin{figure}[ht!]
  \centering
    \includegraphics[width=.59\textwidth]{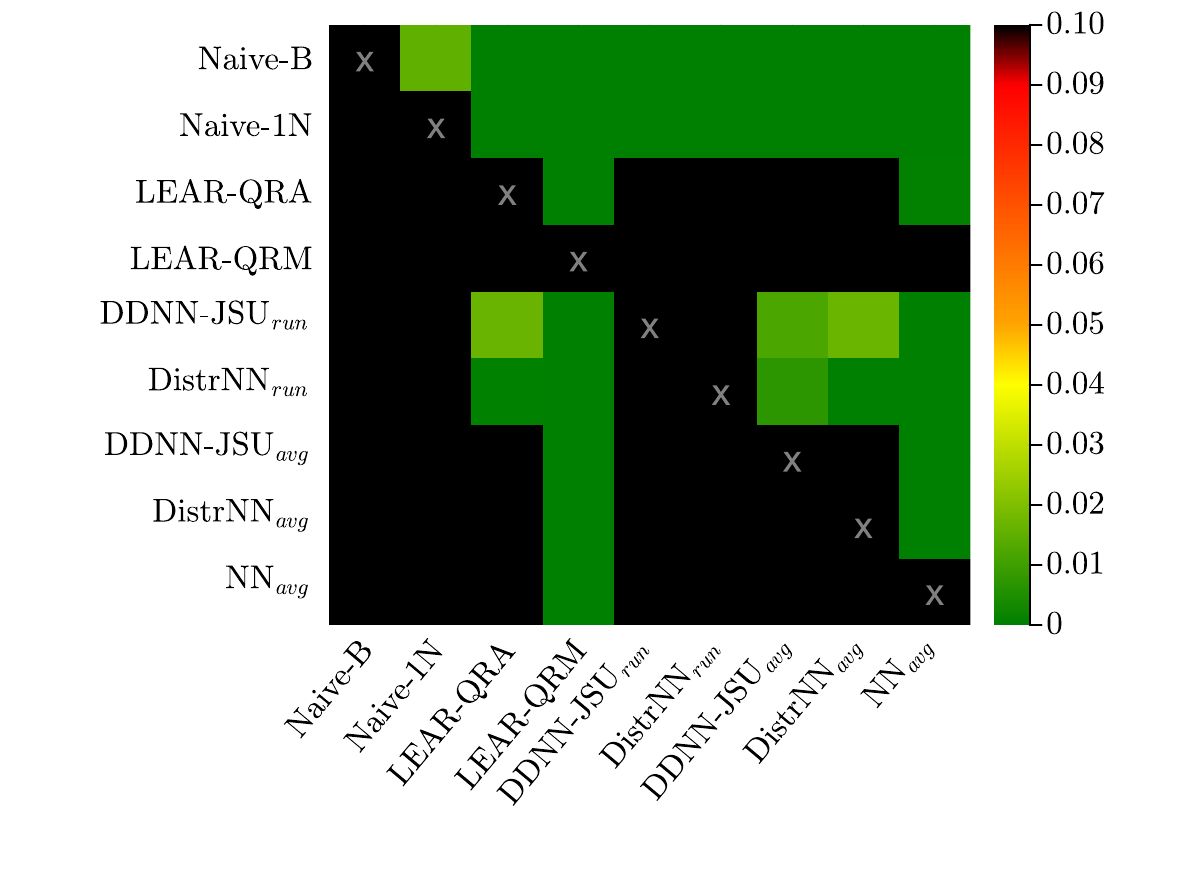}
  \caption{Pairwise DM test $p$-values obtained for the CRPSs aggregated over extreme-tailed quantiles ($1,\ldots,10$ and $90,\ldots,99$) for 2020--2023. The null hypothesis $\mathcal{H}_0$ states that the model on the \textit{y-axis} has a lower predictive loss than the model on the \textit{x-axis} does. Each cell shows the $p$-value produced for the corresponding model comparison. The colors of the cells (nonblack) indicate that the model on the \textit{x-axis} performs significantly better than does the model on the \textit{y-axis} at the 10\% significance level. The black cells indicate $p$-values that are greater than 10\%, implying that no statistically significant performance differences are observed.}
  \label{fig:dm-crps-matrix-tails}
\end{figure}

Presenting the DM test results produced for extreme tails, Figure~\ref{fig:dm-crps-matrix-tails} shows that DistrNN${_\text{avg}}$ and DDNN-JSU${_\text{avg}}$ demonstrate competitive performance in these regions, but their comparisons remain statistically insignificant when they are evaluated over the entire OOS period. In contrast, NN${_\text{avg}}$ demonstrates statistical results against all competitors, except LEAR-QRM, suggesting its strength in terms of capturing extreme quantiles.

Overall, these results are consistent with the findings of \citet{lipiecki2024postprocessing} in that quantile regression-based methods perform better in periods exhibiting structural change and large shifts in distributions, where plain vanilla DistrNNs may struggle.\footnote{In addition, LEAR-QRA and LEAR-QRM are sensitive to the LASSO, the quantile regression estimation algorithm, and the convergence criteria, and the averaging of multiple models improves these models.} The documented results indicate that methods requiring large training datasets (in-sample) to obtain optimized models are limited in that their adaptability to sudden changes is not sufficiently fast. On the other hand, when the distributions of the periods are stable to some extent, the neural network models are more accurate than the linear models are.

\begin{figure}[ht]
  \centering
  \includegraphics[width=0.95\textwidth]{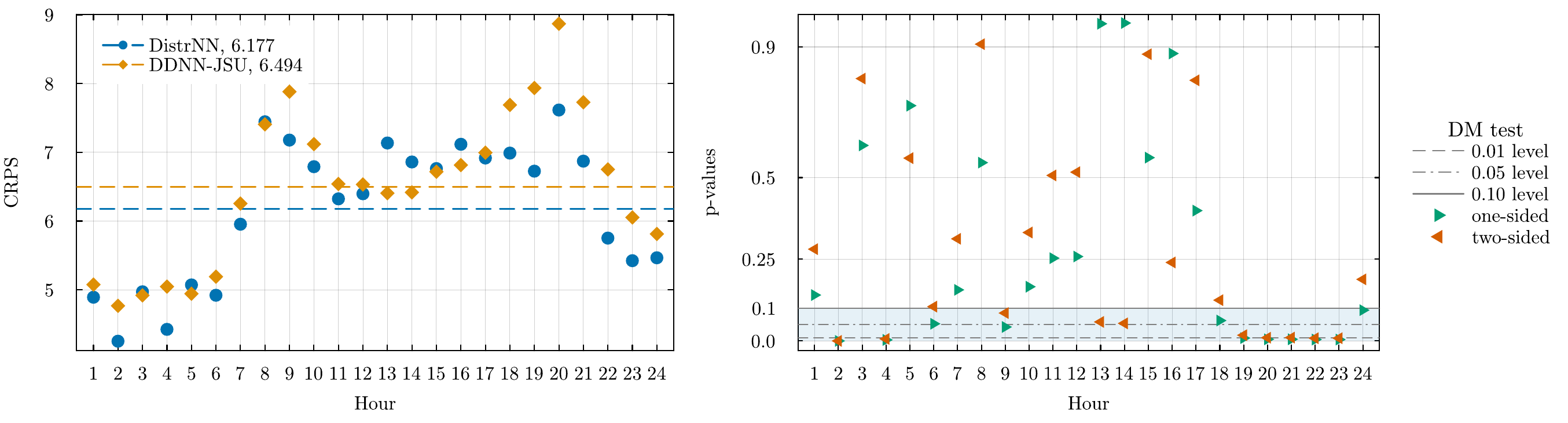}
  \caption{Overall performance, disaggregated by hour. \textit{Left:} CRPS values produced across hourly intervals. \textit{Right:} $p$-values derived from DM tests, disaggregated by hour. For the one-sided test, the null hypothesis $\mathcal{H}_0$ states that the loss of DDNN-JSU$_\text{run}$ is lower than that of DistrNN$_\text{run}$. For the two-sided test, the null hypothesis $\mathcal{H}_0$ states that no significant loss difference is observed between the two models.}
  \label{fig:dm-distr-ddnn-hours}
\end{figure}
In addition, Figure~\ref{fig:dm-distr-ddnn-hours} presents a disaggregated analysis of the performance achieved by the different models by hour, with the left panel showing the CRPS values and the right panel displaying the $p$-values derived from the DM test. This hourly decomposition scheme offers a more granular evaluation of distributional forecasting accuracy. Building on the earlier results, which showed that the DistrNN provides good accuracy in terms of the total daily CRPS, the figure reveals that the DistrNN delivers statistically better forecasts than DDNN-JSU does in approximately half of the hours at the 10\% significance level. Moreover, the three hours with nonsignificant $p$-values fall just short of this threshold, whereas only a few hours indicate relative underperformance, with $p$-values approaching or exceeding 90\%. This suggests that the DistrNN has the potential to capture intraday price fluctuations, which are critical in electricity markets where the dynamics of supply and demand vary significantly throughout the day. 

\begin{figure}[ht]
  \centering
  \includegraphics[width=0.95\textwidth]{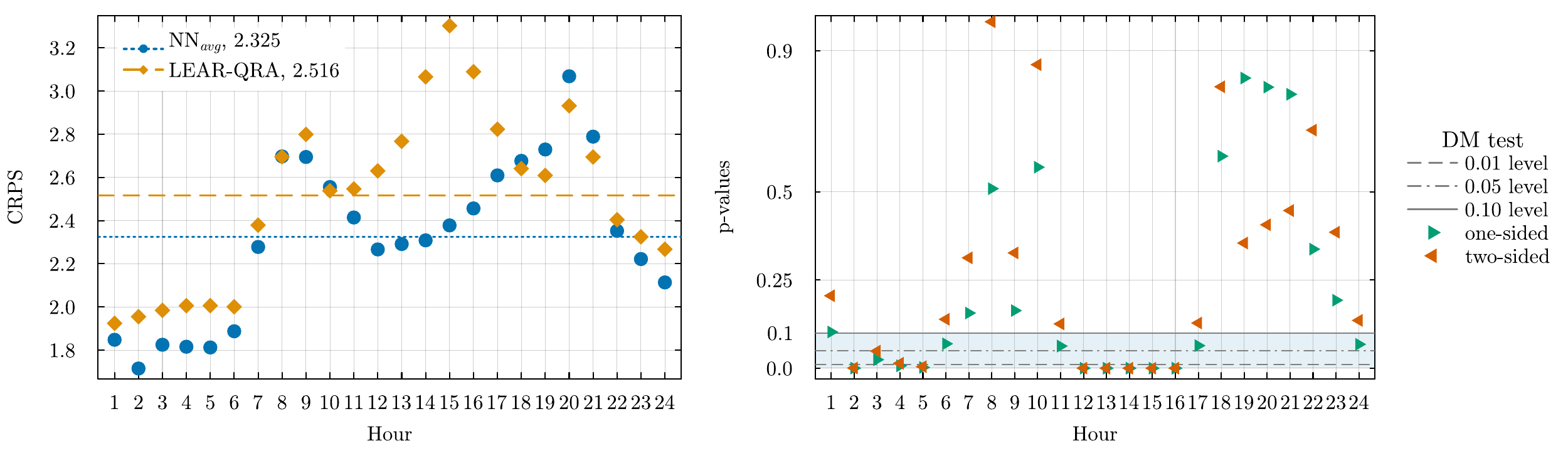}
  \includegraphics[width=0.95\textwidth]{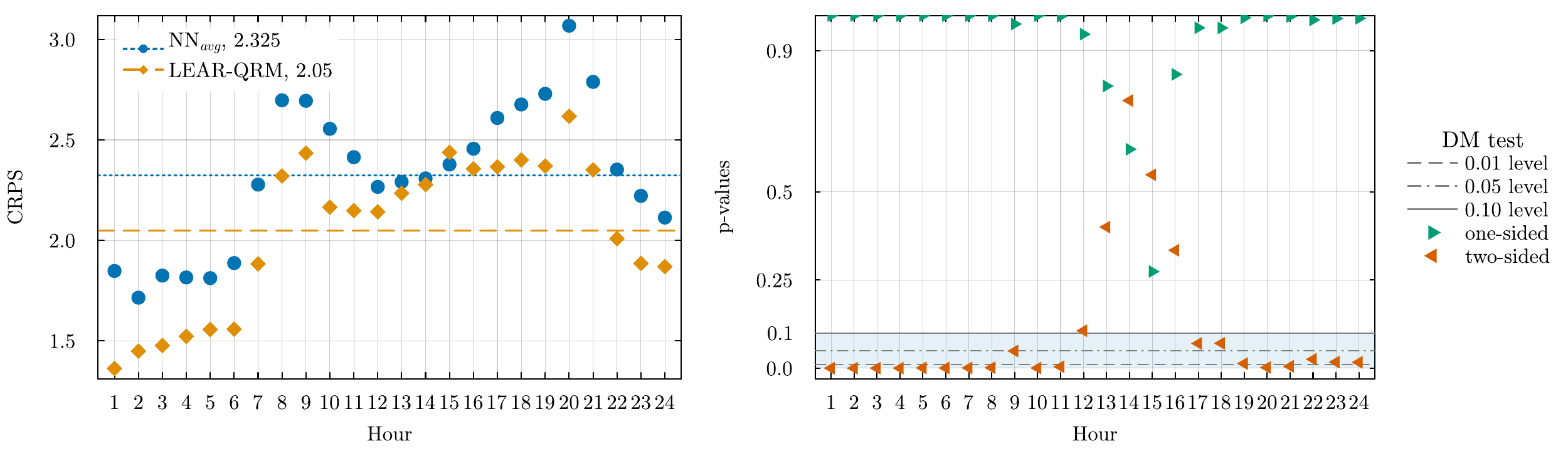}
  \caption{Performance achieved in the extreme tails, disaggregated by hour. \textit{Left:} CRPS values achieved across hourly intervals. \textit{Right:} $p$-values derived from DM tests, disaggregated by hour. For the one-sided test, the null hypothesis $\mathcal{H}_0$ states that the loss of LEAR-QRA (top panel) or LEAR-QRM (bottom panel) is lower than that of NN$_\text{avg}$, where NN$_\text{avg}$ is the average performance of DDNN-JSU$_\text{run}$ or DistrNN$_\text{run}$, respectively. For the two-sided test, the null hypothesis $\mathcal{H}_0$ states that no significant loss difference is observed between the two models.}
  \label{fig:crps-dm-qra-nn-hours-tails}
\end{figure}
Figure~\ref{fig:crps-dm-qra-nn-hours-tails} shows the performance of the distributional forecasts in terms of capturing extreme tail events over different hours of the day, comparing NN$_\text{avg}$ with the LEAR-QRA and LEAR-QRM models. NN$_\text{avg}$ consistently has lower CRPS values than LEAR-QRA does over most hours, even during the peak hours, where the CRPS values generally increase for all models, suggesting greater forecasting difficulty during these periods. The right panel provides a statistical evaluation of the observed differences. 
In contrast, LEAR-QRM of \citet{lipiecki2024postprocessing} exhibits better performance over most of the hours in the extreme tails, except at midday, where the values are not significantly different. For both the top and bottom panels, the results show that significant differences occur primarily in the early morning hours and that NN$_\text{avg}$ favors the midday hours. However, during certain periods, the differences remain statistically insignificant.

\subsection{Comparison with the existing methods}

Our approach differs from the state-of-the-art model \citep{WeronMarcjasz:2023aa} in the literature in terms of its model architecture, computational cost and predictive stability. \citep{WeronMarcjasz:2023aa} employed a larger network with an extensive hyperparameter search process, which can manually activate and deactivate the input features while incorporating regularization techniques such as dropout and $L_2$-norm regularization.

In contrast, our approach uses a narrower network architecture, with a maximum of 256 neurons (compared with 1024), while incorporating a broader set of input features. We also constrain the hyperparameter search space and use AdamW optimization, which inherently mimics $L_2$-norm regularizations. In addition, while \citet{WeronMarcjasz:2023aa} explored 2048 hyperparameter configurations, our method is optimized over a significantly smaller space containing 40 configurations, resulting in a more computationally efficient framework while maintaining competitive prediction accuracy. Comparing the computational costs of these approaches is not straightforward, although the time requirements of both methods are similar for forward rolling schemes. The architecture of the DistrNN in this paper is narrower and has fewer values in its output layer. However, the model must be estimated separately for each of the 24 hours in a day. In contrast, DDNN-JSU \citep{WeronMarcjasz:2023aa} estimates all four parameters of Johnson's SU simultaneously for each hour, resulting in a total of 96 distribution parameters estimated at once.

Our CRPS scores differ from those reported in \citet{WeronMarcjasz:2023aa} and \citet{lipiecki2024postprocessing}. As noted in \citet{lipiecki2024postprocessing}, data changes affected their reported values, with \citet{WeronMarcjasz:2023aa} obtaining a value of 1.304 and \citet{lipiecki2024postprocessing} reporting a result of 1.342 for the period labeled ``2020''. When we replicate the same experimental setup and utilize the same dataset, we obtain a CRPS of 1.329. Given the inherent stochasticity of neural networks, which the authors estimate can introduce a variability level of approximately $\pm 0.1$, this difference is within the expected limits, and a similar situation occurs for the other test periods. In addition, the differences between the LEAR-QRA and LEAR-QRM results, as discussed in \citet{lipiecki2024postprocessing}, are likely due to variations in the LASSO estimation approach. We report the results of LEAR-QRA according to \citet{WeronMarcjasz:2023aa}, where we obtain a value of 1.649 (versus 1.575 in the original report), and for LEAR-QRM, the result is 1.662. These results are higher than the 1.352 value achieved by LEAR-QRM in \citet{lipiecki2024postprocessing}. We see a benefit in reporting both types of results, as well as for future comparisons involving new subperiods of the data sample.\footnote{We report both the models and results that are provided in the literature, as doing so also shows the progress made in the probabilistic electricity price forecasting field.}

In addition, \citet{berrisch2023multivariate} further improved the results of \citet{WeronMarcjasz:2023aa} with their technique of conducting CRPS learning on already-provided OOS results derived from different models. They described how to average such results to obtain a more accurate ensemble average. We do not compete with these results, as the technique can be applied equally well to our DistrNN results.

Furthermore, we do not restrict the reader to taking these results as definitive or to using our approach only as a feedforward neural network. Potential accuracy and performance benefits may be derived if the distributional network is a recurrent, convolutional, temporal-attentional, or other type of network, particularly in cases with trends and structural shifts in data, as occurred in the German market in 2021 and 2022. This also opens up space for further analyses considering parsimony and computational costs. Figure~\ref{fig:hyperopt-app} and Tables~\ref{tab:node_activations} and~\ref{tab:params_hpo} in the appendix show a comparison involving the chosen hyperparameter sets.

\subsection{Software and computational time}

The use of software, particularly in econometrics, has progressed rapidly in recent years. We provide \textsc{Julia} \citep{bezanson2012julia} code with the modelling and estimation procedures, which also serves as an application of a programming environment in a language other \textsc{python} or \textsc{R}. This implementation uses the Flux.jl package \citep{Flux2018}, and the results can be replicated using the examples published at \url{https://github.com/luboshanus/DistrNNEnergy.jl}, which may make it easier for those using Julia to predict (energy) time series.\footnote{The code uses several Julia packages provided in the Project.toml file.}

The full estimation process of the DistrNN involves hyperparameter optimization and a forward rolling window scheme implemented over 24 hours, 1649 OOS observations, 40 hyperparameter sets, 5 folds, and 4 ensembles, with 1500 epochs and a minibatch size of 32 or 64. This process requires the estimation of 4800 (24*40*5) and 384 (24*4*4) 1500-epoch training models and 158304 (24*1649*4) 500-epoch neural network updates. Consequently, obtaining an OOS prediction can take approximately 10--17 hours, depending significantly on the number of nodes contained in the selected neural network used in the rolling scheme. To manage these computations, we distribute the hyperparameter optimization and rolling estimation tasks across 40 CPU cores.\footnote{We use 40 CPU cores of the 48-core Intel(R) Xeon(R) Gold 6542Y processor.}
The complete estimation process of LEAR-QRA(QRM) in Julia takes approximately 15 minutes when distributed over 30 CPU cores.\footnote{We rewrite parts of the authors' open-access toolboxes \citep{Lago:2021codeEPF,WeronMarcjasz:2023aa} in Julia.}

% _________
\section{Conclusion}

A novel machine learning approach for probabilistically forecasting hourly day-ahead electricity prices is proposed in this paper. Compared with the state-of-the-art frameworks in the probabilistic electricity price forecasting literature, our model provides more accurate forecasts that uncover new valuable information in the input data, especially by using the dynamics of probabilities itself. This is mainly because it does not rely on restrictive model assumptions and allows for non-Gaussian, heavy-tailed data and their nonlinear interactions. By relaxing the assumption regarding the distribution family of the given time series, our distributional neural network fully explores the data. We also provide an efficient computational package that can be used by researchers.

\onehalfspacing
% \linespread{1}
\begingroup
\setlength{\bibsep}{0pt}
\setlength{\bibhang}{1.0em}

\bibliographystyle{chicago}
\bibliography{bibs/BiblioDistributions}
\endgroup

% ~~~~~~~~~~~~~~~~~~~~~~ Tables and Figures ~~~~~~~~~~~~~~~~~~~~~~~~ %
% ~~~~~~~~~~~~~~~~~~~~~~~~~~~~~~~~~~~~~~~~~~~~~~~~~~~~~~~~~~~~~~~~~~~%
\newpage

\setcounter{section}{0}
\setcounter{equation}{0}
\setcounter{figure}{0}
\setcounter{table}{0}

\def\thesection{\Alph{section}}
\def\thesubsection{\thesection.\arabic{subsection}}
\def\thesubsubsection{\thesubsection.\arabic{subsubsection}}
\renewcommand{\theequation}{\Alph{section}.\arabic{equation}}
\renewcommand{\thetable}{A\arabic{table}}
\renewcommand{\thefigure}{A\arabic{figure}}

\begin{center}
    \Large \textbf{Appendix for} 
\end{center}
\begin{center}
    \Large
    ``Learning the Probability Distributions of Day-ahead Electricity Prices''
\end{center}

\section{Additional tables and figures}
\label{app:tables}

% \clearpage
\subsection{Hyperoptimization results}
\begin{figure}[ht]
  \centering
  \includegraphics[width=0.99\textwidth]{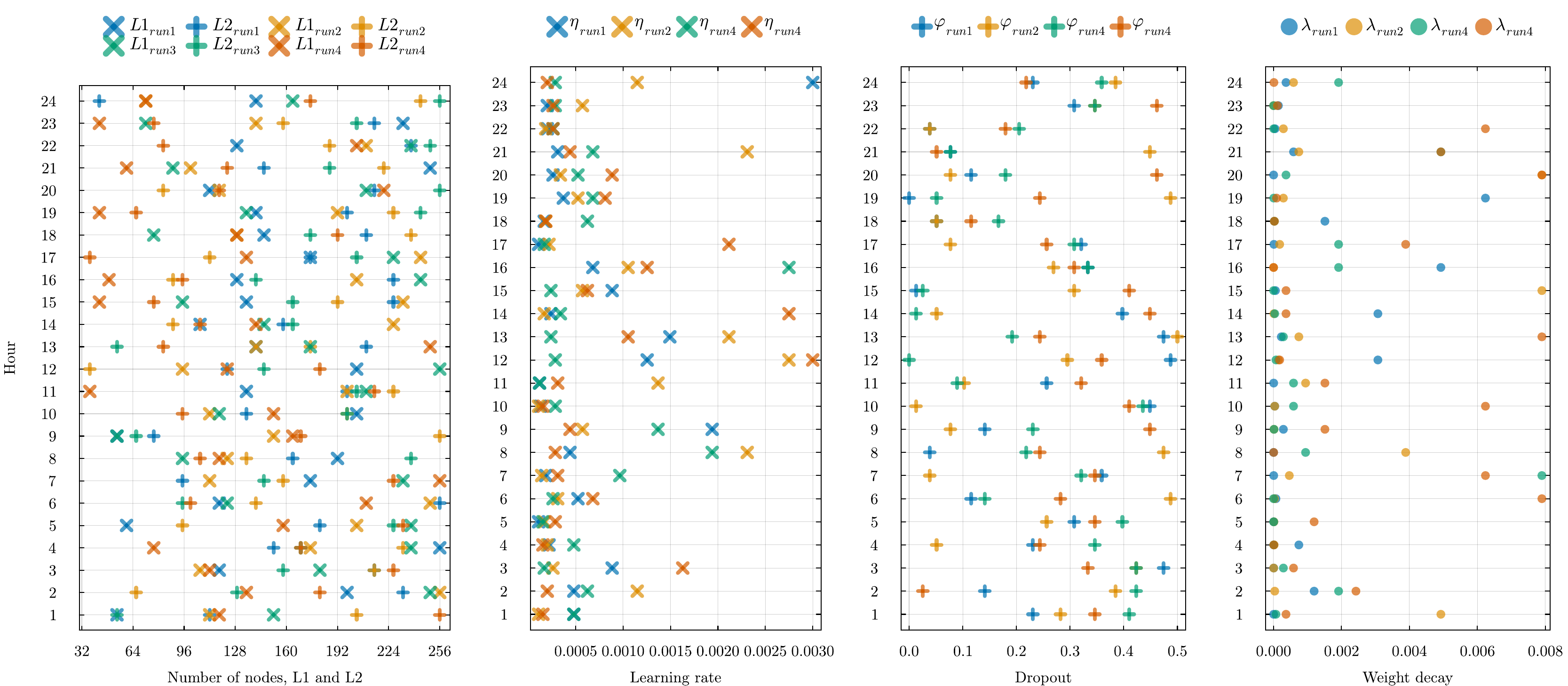}
  \caption{Hyperoptimization results of the parameter sets selected for all hours. The figures show the values of the parameters that make up the best sets used to relearn/recalibrate the neural networks for each hour. Together with these parameters, the hyperoptimization scheme selects the activation function, and all selected hyperparameters are displayed in Tables~\ref{tab:node_activations} and~\ref{tab:params_hpo}.}
  \label{fig:hyperopt-app}
\end{figure}

Here, we provide figures and tables depicting the selected hyperparameters for all hours, which are estimated via the DistrNN; see Figure~\ref{fig:hyperopt-app} and Tables~\ref{tab:node_activations} and~\ref{tab:params_hpo}.

\begin{table}[htbp]
  \centering
  \caption{The number of neurons selected in each layer}
  {\fontsize{7}{9}\selectfont
  \setlength\tabcolsep{5pt}
\begin{tabular}{rrrrrrrrrrrrrrllrllrllrll}
\toprule
  &    & \multicolumn{11}{c}{Number of nodes}                 &    & \multicolumn{11}{c}{Activation functions} \\
\cmidrule{3-4}\cmidrule{6-7}\cmidrule{9-10}\cmidrule{12-13}\cmidrule{15-16}\cmidrule{18-19}\cmidrule{21-22}\cmidrule{24-25}   &    & \multicolumn{2}{c}{run$_1$} &    & \multicolumn{2}{c}{run$_2$} &    & \multicolumn{2}{c}{run$_3$} &    & \multicolumn{2}{c}{run$_4$} &    & \multicolumn{2}{c}{run$_1$} &    & \multicolumn{2}{c}{run$_2$} &    & \multicolumn{2}{c}{run$_3$} &    & \multicolumn{2}{c}{run$_4$} \\
\multicolumn{1}{l}{hour} &    & \multicolumn{1}{l}{L1} & \multicolumn{1}{l}{L2} &    & \multicolumn{1}{l}{L1} & \multicolumn{1}{l}{L2} &    & \multicolumn{1}{l}{L1} & \multicolumn{1}{l}{L2} &    & \multicolumn{1}{l}{L1} & \multicolumn{1}{l}{L2} &    & L1 & L2 &    & L1 & L2 &    & L1 & L2 &    & L1 & L2 \\
\cmidrule{3-4}\cmidrule{6-7}\cmidrule{9-10}\cmidrule{12-13}\cmidrule{15-16}\cmidrule{18-19}\cmidrule{21-22}\cmidrule{24-25}1  &    & 54 & 112 &    & 118 & 256 &    & 152 & 54 &    & 95 & 95 &    & soft$^+$ & ELU&    & soft$^+$ & soft$^+$ &    & soft$^+$ & ELU&    & tanh & soft$^+$ \\
2  &    & 198 & 233 &    & 135 & 181 &    & 250 & 129 &    & 66 & 152 &    & soft$^+$ & ELU&    & soft$^+$ & ELU &    & soft$^+$ & ELU&    & soft$^+$ & soft$^+$ \\
3  &    & 118 & 215 &    & 112 & 227 &    & 181 & 158 &    & 256 & 164 &    & soft$^+$ & ReLU&    & ReLU & soft$^+$ &    & soft$^+$ & soft$^+$ &    & soft$^+$ & soft$^+$ \\
4  &    & 256 & 152 &    & 77 & 169 &    & 238 & 169 &    & 95 & 135 &    & soft$^+$ & soft$^+$ &    & ELU& soft$^+$ &    & soft$^+$ & ReLU&    & tanh & soft$^+$ \\
5  &    & 60 & 181 &    & 158 & 233 &    & 238 & 227 &    & 112 & 106 &    & soft$^+$ & ELU&    & soft$^+$ & soft$^+$ &    & soft$^+$ & soft$^+$ &    & ELU& soft$^+$ \\
6  &    & 118 & 256 &    & 210 & 100 &    & 123 & 95 &    & 112 & 152 &    & ReLU& soft$^+$ &    & soft$^+$ & soft$^+$ &    & soft$^+$ & ELU&    & ELU& soft$^+$ \\
7  &    & 175 & 95 &    & 256 & 227 &    & 233 & 146 &    & 118 & 118 &    & soft$^+$ & soft$^+$ &    & ReLU& soft$^+$ &    & soft$^+$ & soft$^+$ &    & ELU& soft$^+$ \\
8  &    & 192 & 164 &    & 118 & 106 &    & 95 & 238 &    & 37 & 89 &    & ReLU& ReLU&    & ReLU& ELU&    & ReLU& ReLU&    & ELU& soft$^+$ \\
9  &    & 54 & 77 &    & 164 & 169 &    & 54 & 66 &    & 210 & 32 &    & soft$^+$ & ELU&    & soft$^+$ & soft$^+$ &    & tanh & ReLU&    & sigmoid & ReLU\\
10 &    & 204 & 135 &    & 152 & 95 &    & 118 & 198 &    & 256 & 100 &    & soft$^+$ & soft$^+$ &    & soft$^+$ & soft$^+$ &    & soft$^+$ & ELU&    & soft$^+$ & soft$^+$ \\
11 &    & 135 & 198 &    & 37 & 215 &    & 210 & 204 &    & 32 & 164 &    & ELU& soft$^+$ &    & soft$^+$ & soft$^+$ &    & soft$^+$ & ELU&    & ELU& ELU\\
12 &    & 204 & 123 &    & 123 & 181 &    & 256 & 146 &    & 54 & 215 &    & ELU& ReLU&    & ELU& soft$^+$ &    & soft$^+$ & ReLU&    & ELU& ReLU\\
13 &    & 141 & 210 &    & 250 & 83 &    & 175 & 54 &    & 83 & 77 &    & soft$^+$ & ELU&    & ReLU& soft$^+$ &    & ELU& ELU&    & ELU& ReLU\\
14 &    & 106 & 158 &    & 141 & 106 &    & 146 & 164 &    & 112 & 164 &    & soft$^+$ & ReLU&    & ReLU& soft$^+$ &    & soft$^+$ & ELU&    & ELU& soft$^+$ \\
15 &    & 135 & 227 &    & 43 & 77 &    & 95 & 164 &    & 187 & 210 &    & soft$^+$ & ELU&    & ReLU& soft$^+$ &    & ReLU& soft$^+$ &    & soft$^+$ & soft$^+$ \\
16 &    & 129 & 227 &    & 49 & 95 &    & 244 & 141 &    & 83 & 164 &    & soft$^+$ & ReLU&    & ReLU& soft$^+$ &    & ELU& ReLU&    & sigmoid & soft$^+$ \\
17 &    & 175 & 175 &    & 135 & 37 &    & 227 & 204 &    & 215 & 175 &    & soft$^+$ & soft$^+$ &    & ReLU& soft$^+$ &    & soft$^+$ & ELU&    & soft$^+$ & soft$^+$ \\
18 &    & 146 & 210 &    & 129 & 192 &    & 77 & 175 &    & 60 & 95 &    & soft$^+$ & ELU&    & soft$^+$ & ELU&    & soft$^+$ & ReLU&    & ReLU& soft$^+$ \\
19 &    & 141 & 198 &    & 43 & 66 &    & 135 & 244 &    & 187 & 89 &    & soft$^+$ & ReLU&    & soft$^+$ & soft$^+$ &    & soft$^+$ & ELU&    & tanh & soft$^+$ \\
20 &    & 112 & 215 &    & 221 & 118 &    & 210 & 256 &    & 158 & 210 &    & soft$^+$ & ELU&    & ReLU& soft$^+$ &    & ReLU& ReLU&    & soft$^+$ & soft$^+$ \\
21 &    & 250 & 146 &    & 60 & 123 &    & 89 & 187 &    & 152 & 89 &    & soft$^+$ & ELU&    & soft$^+$ & soft$^+$ &    & ReLU& soft$^+$ &    & soft$^+$ & soft$^+$ \\
22 &    & 129 & 238 &    & 204 & 83 &    & 238 & 250 &    & 175 & 164 &    & soft$^+$ & ELU&    & ReLU& soft$^+$ &    & ReLU& tanh &    & ELU& soft$^+$ \\
23 &    & 233 & 215 &    & 43 & 77 &    & 72 & 204 &    & 37 & 192 &    & soft$^+$ & soft$^+$ &    & ReLU& ReLU&    & ReLU& soft$^+$ &    & ELU& soft$^+$ \\
24 &    & 141 & 43 &    & 72 & 175 &    & 164 & 256 &    & 158 & 164 &    & ReLU& ELU&    & soft$^+$ & soft$^+$ &    & soft$^+$ & soft$^+$ &    & tanh & soft$^+$ \\
\cmidrule{3-4}\cmidrule{6-7}\cmidrule{9-10}\cmidrule{12-13}\cmidrule{15-16}\cmidrule{18-19}\cmidrule{21-22}\cmidrule{24-25}
\bottomrule
\end{tabular}%
    }
  \label{tab:node_activations}%
  \caption*{The table describes the selected numbers of nodes for the two hidden layers for each hour and each run of the DistrNN. Furthermore, the right part of the table shows activation functions chosen after each of the two layers. $\text{soft}^+$ denotes the softplus layer.}
\end{table}%

\begin{table}[htbp]
  \centering
  \caption{The parameters selected for every hour}
    {\fontsize{8}{10}\selectfont
    \setlength\tabcolsep{5pt}
\begin{tabular}{rrrrrrrrrrrrrrrr}
\cmidrule{3-6}\cmidrule{8-11}\cmidrule{13-16}
 &    & \multicolumn{4}{c}{Learning rate, $\eta$} &    & \multicolumn{4}{c}{Weight decay rate,  $\lambda$} &    & \multicolumn{4}{c}{Dropout rate, $\phi$} \\
\multicolumn{1}{l}{hour} &    & \multicolumn{1}{l}{run$_1$} & \multicolumn{1}{l}{run$_2$} & \multicolumn{1}{l}{run$_3$} & \multicolumn{1}{l}{run$_4$} &    & \multicolumn{1}{l}{run$_1$} & \multicolumn{1}{l}{run$_2$} & \multicolumn{1}{l}{run$_3$} & \multicolumn{1}{l}{run$_4$} &    & \multicolumn{1}{l}{run$_1$} & \multicolumn{1}{l}{run$_2$} & \multicolumn{1}{l}{run$_3$} & \multicolumn{1}{l}{run$_4$} \\
\cmidrule{3-6}\cmidrule{8-11}\cmidrule{13-16}1  &    & 0.00048 & 0.00011 & 0.00048 & 0.00015 &    & 0.00000 & 0.00492 & 0.00007 & 0.00037 &    & 0.23 & 0.28 & 0.41 & 0.35 \\
2  &    & 0.00048 & 0.00115 & 0.00062 & 0.00020 &    & 0.00119 & 0.00003 & 0.00191 & 0.00242 &    & 0.14 & 0.38 & 0.42 & 0.03 \\
3  &    & 0.00088 & 0.00026 & 0.00017 & 0.00163 &    & 0.00001 & 0.00000 & 0.00029 & 0.00059 &    & 0.47 & 0.42 & 0.42 & 0.33 \\
4  &    & 0.00022 & 0.00020 & 0.00048 & 0.00015 &    & 0.00074 & 0.00002 & 0.00001 & 0.00001 &    & 0.23 & 0.05 & 0.35 & 0.24 \\
5  &    & 0.00011 & 0.00017 & 0.00015 & 0.00028 &    & 0.00000 & 0.00001 & 0.00001 & 0.00119 &    & 0.31 & 0.26 & 0.40 & 0.35 \\
6  &    & 0.00052 & 0.00031 & 0.00026 & 0.00068 &    & 0.00007 & 0.00003 & 0.00000 & 0.00790 &    & 0.12 & 0.49 & 0.14 & 0.28 \\
7  &    & 0.00018 & 0.00014 & 0.00097 & 0.00031 &    & 0.00001 & 0.00046 & 0.00790 & 0.00624 &    & 0.36 & 0.04 & 0.32 & 0.35 \\
8  &    & 0.00044 & 0.00231 & 0.00194 & 0.00028 &    & 0.00000 & 0.00389 & 0.00094 & 0.00000 &    & 0.04 & 0.47 & 0.22 & 0.24 \\
9  &    & 0.00194 & 0.00057 & 0.00137 & 0.00044 &    & 0.00029 & 0.00001 & 0.00001 & 0.00151 &    & 0.14 & 0.08 & 0.23 & 0.45 \\
10 &    & 0.00013 & 0.00011 & 0.00028 & 0.00015 &    & 0.00003 & 0.00003 & 0.00059 & 0.00624 &    & 0.45 & 0.01 & 0.44 & 0.41 \\
11 &    & 0.00012 & 0.00137 & 0.00012 & 0.00031 &    & 0.00000 & 0.00094 & 0.00059 & 0.00151 &    & 0.26 & 0.10 & 0.09 & 0.32 \\
12 &    & 0.00125 & 0.00275 & 0.00028 & 0.00300 &    & 0.00307 & 0.00014 & 0.00007 & 0.00018 &    & 0.49 & 0.29 & 0.00 & 0.36 \\
13 &    & 0.00149 & 0.00212 & 0.00024 & 0.00105 &    & 0.00023 & 0.00074 & 0.00029 & 0.00790 &    & 0.47 & 0.50 & 0.19 & 0.24 \\
14 &    & 0.00024 & 0.00017 & 0.00034 & 0.00275 &    & 0.00307 & 0.00001 & 0.00003 & 0.00037 &    & 0.40 & 0.05 & 0.01 & 0.45 \\
15 &    & 0.00088 & 0.00057 & 0.00024 & 0.00062 &    & 0.00006 & 0.00790 & 0.00000 & 0.00037 &    & 0.01 & 0.31 & 0.03 & 0.41 \\
16 &    & 0.00068 & 0.00105 & 0.00275 & 0.00125 &    & 0.00492 & 0.00000 & 0.00191 & 0.00000 &    & 0.33 & 0.27 & 0.33 & 0.31 \\
17 &    & 0.00011 & 0.00022 & 0.00017 & 0.00212 &    & 0.00001 & 0.00018 & 0.00191 & 0.00389 &    & 0.32 & 0.08 & 0.31 & 0.26 \\
18 &    & 0.00017 & 0.00018 & 0.00062 & 0.00018 &    & 0.00151 & 0.00003 & 0.00002 & 0.00002 &    & 0.05 & 0.05 & 0.17 & 0.12 \\
19 &    & 0.00037 & 0.00052 & 0.00068 & 0.00081 &    & 0.00624 & 0.00029 & 0.00001 & 0.00009 &    & 0.00 & 0.49 & 0.05 & 0.24 \\
20 &    & 0.00026 & 0.00034 & 0.00052 & 0.00088 &    & 0.00000 & 0.00790 & 0.00037 & 0.00790 &    & 0.12 & 0.08 & 0.18 & 0.46 \\
21 &    & 0.00031 & 0.00231 & 0.00068 & 0.00044 &    & 0.00059 & 0.00074 & 0.00492 & 0.00492 &    & 0.08 & 0.45 & 0.08 & 0.05 \\
22 &    & 0.00020 & 0.00018 & 0.00026 & 0.00026 &    & 0.00004 & 0.00029 & 0.00001 & 0.00624 &    & 0.04 & 0.04 & 0.21 & 0.18 \\
23 &    & 0.00020 & 0.00057 & 0.00028 & 0.00026 &    & 0.00014 & 0.00000 & 0.00000 & 0.00011 &    & 0.31 & 0.35 & 0.35 & 0.46 \\
24 &    & 0.00300 & 0.00115 & 0.00028 & 0.00020 &    & 0.00037 & 0.00059 & 0.00191 & 0.00001 &    & 0.23 & 0.38 & 0.36 & 0.22 \\
\cmidrule{3-6}\cmidrule{8-11}\cmidrule{13-16}
\bottomrule
\end{tabular}%
    }
  \label{tab:params_hpo}%
  \caption*{This table provides the parameters selected for each hour and each run of the DistrNN.}
\end{table}%

% _______
\clearpage
\section{CDF interpolation}
\label{app:interpolation}
The Fritsch--Carlson monotonic cubic interpolation scheme \citep{fritsch1980interpolation}
provides a monotonically increasing CDF with a range of $[0,1]$ when applied to CDF estimates on a finite grid.

Suppose that we have a CDF $F(y)$ defined at points $(y_k,F(y_k))$ for $k=1,\dots ,K,$ where $F(y_0)=0$ and $F(y_K)=1$.
We assume that $y_k<y_{k+1}$ and $F(y_k)<F(y_{k+1})$ for all $k=0,\dots ,K-1,$, which is warranted by the continuity of returns and the construction of the estimated distribution. First, we compute the slopes of the secant lines as $\Delta_k =(F(y_{k+1})-F(y_k)))/(y_{k+1}-y_k)$ for $k=1,\dots,K-1;$ then, the tangents at every data point are determined as $m_1 = \Delta_1$, $m_k = \frac{1}{2}(\Delta_{k-1}+\Delta_k)$ for $k=2,\dots,K-1$, and $m_K = \Delta_{K-1}.$
Let $\alpha_k = m_k/\Delta_k$ and $\beta_k = m_{k+1}/\Delta_k$ for $k=1,\dots,K-1$.
If $\alpha_k^2 + \beta_k^2 > 9$ for some $k=1,\dots,K-1,$, then we set $m_k = \alpha_k \alpha_k \Delta_k$ and $m_{k+1} = \alpha_k \beta_k \Delta_k,$ with $\alpha_k = 3(\alpha_k^2 + \beta_k^2)^{-1/2}$.
Finally, the cubic Hermite spline method is applied: for any $y\in [y_k,y_{k+1}]$ for some $k=0,\dots,K-1,$
we evaluate $F(y)$ as
$$F(y) = (2t^3-3t^2+1)F(y_k)  + (t^3-2t^2+t)h y_{k}  + (-2t^3+3t^2) F(y_{k+1}) + (t^3-t^2)h m_{k+1},$$ where $h=y_{k+1}-y_{k}$ and $t=(y-y_{k})/h.$

\end{document}